\documentclass[a4paper,11pt]{article}
\pdfoutput=1
\usepackage{jcappub} 
\usepackage{lineno}

\usepackage[T1]{fontenc} 
\usepackage{subcaption}
\usepackage{array}

\usepackage{graphicx}
\usepackage{dcolumn}
\usepackage{bm}
\usepackage{xspace}
\usepackage{xcolor}
\usepackage{hyperref}
\hypersetup{colorlinks=true,citecolor=blue,filecolor=blue,urlcolor=blue,}
\usepackage[noabbrev]{cleveref}
\usepackage{orcidlink}

\newcommand{\lya}{Ly$\alpha$\xspace}
\newcommand{\lyb}{Ly$\beta$\xspace}

\newcommand{\ion}[2]{#1\thinspace{}#2\xspace}

\title{Validation of the DESI 2024 Lyman Alpha Forest BAL Masking Strategy}

\author[1,2,3]{{P.~Martini}\orcidlink{0000-0002-4279-4182},}
\author[2,1,3]{{A.~Cuceu}\orcidlink{0000-0002-2169-0595},}
\author[2,3]{{L.~Ennesser}\orcidlink{0000-0002-9501-8769},}
\author[4,5]{{A.~Brodzeller}\orcidlink{0000-0002-8934-0954},}
\author[5]{{J.~Aguilar},}
\author[6]{{S.~Ahlen}\orcidlink{0000-0001-6098-7247},}
\author[7]{{D.~Brooks},}
\author[5]{{T.~Claybaugh},}
\author[5]{{R.~de Belsunce}\orcidlink{0000-0003-3660-4028},}
\author[8]{{A.~de la Macorra}\orcidlink{0000-0002-1769-1640},}
\author[9]{{Arjun~Dey}\orcidlink{0000-0002-4928-4003},}
\author[7]{{P.~Doel},}
\author[10,11]{{J.~E.~Forero-Romero}\orcidlink{0000-0002-2890-3725},}
\author[12,13,14]{{E.~Gaztañaga},}
\author[5]{{S.~Gontcho A Gontcho}\orcidlink{0000-0003-3142-233X},}
\author[5]{{J.~Guy}\orcidlink{0000-0001-9822-6793},}
\author[15]{{H.~K.~Herrera-Alcantar}\orcidlink{0000-0002-9136-9609},}
\author[2,3,16]{{K.~Honscheid},}
\author[2,1,3]{{N.~G.~Kara{\c c}ayl{\i}}\orcidlink{0000-0001-7336-8912},}
\author[5]{{T.~Kisner}\orcidlink{0000-0003-3510-7134},}
\author[5]{{A.~Kremin}\orcidlink{0000-0001-6356-7424},}
\author[5]{{A.~Lambert},}
\author[17]{{L.~Le~Guillou}\orcidlink{0000-0001-7178-8868},}
\author[18,19]{{M.~Manera}\orcidlink{0000-0003-4962-8934},}
\author[9]{{A.~Meisner}\orcidlink{0000-0002-1125-7384},}
\author[20,19]{{R.~Miquel},}
\author[21]{{P.~Montero-Camacho}\orcidlink{0000-0002-6998-6678},}
\author[22]{{J.~Moustakas}\orcidlink{0000-0002-2733-4559},}
\author[15,23]{{G.~Niz}\orcidlink{0000-0002-1544-8946},}
\author[24,5]{{N.~Palanque-Delabrouille}\orcidlink{0000-0003-3188-784X},}
\author[25,26,27]{{W.~J.~Percival}\orcidlink{0000-0002-0644-5727},}
\author[28]{{I.~P\'erez-R\`afols}\orcidlink{0000-0001-6979-0125},}
\author[5,29,30]{{C.~Poppett},}
\author[31]{{F.~Prada}\orcidlink{0000-0001-7145-8674},}
\author[32,24,33]{{C.~Ravoux}\orcidlink{0000-0002-3500-6635},}
\author[34]{{M.~Rezaie}\orcidlink{0000-0001-5589-7116},}
\author[35]{{G.~Rossi},}
\author[36]{{E.~Sanchez}\orcidlink{0000-0002-9646-8198},}
\author[5]{{D.~Schlegel},}
\author[37,38]{{M.~Schubnell},}
\author[39]{{H.~Seo}\orcidlink{0000-0002-6588-3508},}
\author[9]{{D.~Sprayberry},}
\author[24]{{T.~Tan}\orcidlink{0000-0001-8289-1481},}
\author[38]{{G.~Tarl\'{e}}\orcidlink{0000-0003-1704-0781},}
\author[40,41]{{M.~Walther}\orcidlink{0000-0002-1748-3745},}
\author[9]{{B.~A.~Weaver},}
\author[42]{{H.~Zou}\orcidlink{0000-0002-6684-3997},}

\affiliation{Affiliations are in Appendix \ref{sec:affiliations}}

\emailAdd{martini.10@osu.edu}

\abstract{
Broad absorption line quasars (BALs) exhibit blueshifted absorption relative to a number of their prominent broad emission features. These absorption features can contribute to quasar redshift errors and add absorption to the Lyman-$\alpha$ (\lya) forest that is unrelated to large-scale structure. We present a detailed analysis of the impact of BALs on the Baryon Acoustic Oscillation (BAO) results with the \lya forest from the first year of data from the Dark Energy Spectroscopic Instrument (DESI). The baseline strategy for the first year analysis is to mask all pixels associated with all BAL absorption features that fall within the wavelength region used to measure the forest. We explore a range of alternate masking strategies and demonstrate that these changes have minimal impact on the BAO measurements with both DESI data and synthetic data. This includes when we mask the BAL features associated with emission lines outside of the forest region to minimize their contribution to redshift errors. We identify differences in the properties of BALs in the synthetic datasets relative to the observational data, as well as use the synthetic observations to characterize the completeness of the BAL identification algorithm, and demonstrate that incompleteness and differences in the BALs between real and synthetic data also do not impact the BAO results for the \lya forest. 
}

\begin{document}
\maketitle
\flushbottom

\section{Introduction} \label{sec:intro}

The discovery of the accelerating expansion of the universe in the late 1990s conclusively showed the need for an additional component in the standard cosmological model, one that indicated that a key ingredient is missing from our understanding of physics at a fundamental level. The additional component is commonly parameterized as a cosmological constant and referred to as dark energy \cite[e.g. see][for a review]{Weinberg2013}. Over the intervening decades, progressively larger experiments have produced progressively higher precision measurements of cosmological parameters \cite{Planck2018,Dawson2013,Abbott2018,Alam2021}. These experiments have substantially refined our cosmological model, such as that dark energy constitutes about 70\% of the matter-energy density of the universe at the present day, although have not yet led to a conclusive understanding of the nature of this key component of the universe.

The continuing quest to understand the universe, and especially to explore the dark energy component, led to the development of the Dark Energy Spectroscopic Instrument \citep[DESI][]{Levi2013,DESI2016a.Science,DESI2016b.Instr}. DESI aims to measure the cosmic expansion history with unprecedented precision with a spectroscopic survey of approximately 40 million galaxies and quasars in just five years. The goal of DESI is to use the Baryon Acoustic Oscillation (BAO) technique to measure the cosmic expansion history and the geometry of the universe. The survey targets include galaxies and quasars that span from the local universe to beyond redshift $z>3$, and uses BAO measurements at a range of redshifts as a precise and well established method for the measurement of the matter and energy density of the universe. The DESI survey began in May 2021, and the first year of data includes spectra of over 14 million extragalactic spectra. This is several times larger than all previous samples combined. 

The highest redshift measurements from DESI are observations of quasars. Above redshift $z>2.1$, DESI spectra include the \lya forest, a dense thicket of absorption features due to \lya absorption from neutral Hydrogen atoms in the extremely rarefied, highly ionized intergalactic medium (IGM). The distribution of absorption traced by the \lya forest provides information about the matter distribution along the line of sight to each quasar, and thus contains important information that may be used to determine cosmological parameters, such as through measurement of the BAO scale. 

The first measurements of the BAO scale with \lya forest data calculated the \lya forest auto-correlation function \citep{Busca2013,Slosar2013,Kirkby2013} with about 50,000 quasars observed as part of the Baryon Oscillation Spectroscopic Survey \citep[BOSS, see][]{Dawson2013}, which were part of the Ninth Data Release (DR9) of the Sloan Digital Sky Survey \citep[SDSS, see][]{Eisenstein2011,Ahn2012}. These results were quickly followed by measurements based on the cross-correlation between the \lya forest and QSOs \cite{FontRibera2013}. Subsequent data releases from SDSS \cite{Alam2015,Ahumada2020} lead to more precise measurements \cite{Delubac2015,Bautista2017} that culminated in the best \lya measurement to date \cite{dMdB2020} with over 210,000 quasars at $z>2.1$ for measurement of the \lya forest auto-correlation function and over 340,000 quasars for measurement of the forest cross-correlation with quasars. 

The DESI Early Data Release \cite{DESI2023b.KP1.EDR} presented preliminary results on the \lya forest \cite{LyaBAO.EDR.Gordon.2023} and outlined some of the main methodology \cite{Ramirez2024} employed in the \lya analysis. The quasar sample from the first year of observations, which will be part of the future Data Release 1 \cite[DR1,][]{DESI2024.I.DR1}, represents a substantial increase in sample size compared to previous work, with over 450,000 \lya spectra and over 700,000 quasars for measurement of the cross correlation \cite{DESI2024.IV.KP6}. The results from this analysis of DR1 consequently represent the most precise and rigorously tested \lya forest measurements to date. The DESI 2024 Key Paper on the \lya forest \cite{DESI2024.IV.KP6} reports measurements on the expansion $H(z_{eff})$ at $z_{eff} = 2.33$ with better than 2\% precision, and the transverse comoving distance $D_M(z_{eff})$ with 2.4\% precision. Several companion papers present supporting analysis details, including a thorough study of the analysis pipeline with synthetic data \cite{KP6s6-Cuceu} and a detailed investigation of the impact of instrumental systematics \cite{KP6s5-Guy}. 

This paper also supports the DESI DR1 \lya results \cite{DESI2024.IV.KP6} with a thorough investigation of the impact of systematics related to Broad Absorption Line (BAL) quasars on the BAO measurements. BAL quasars exhibit blueshifted absorption relative to a number of the broad emission features that are characteristic of quasars, including some that fall within the wavelength range of the \lya forest. The BAL features consequently absorb some of the quasar continuum in the forest region, and typically it is not possible to distinguish BAL features from absorption by neutral Hydrogen in the IGM. Furthermore, this absorption is present in 10--30\% of the quasar population \cite{Foltz1990,Trump2006,Paris2017}, depending on spectroscopic data quality and the quasar selection algorithm. The BAL fraction ranged from 12--20\% in the DESI EDR quasar sample \cite{Filbert2023}.  

DESI employs a strategy to mitigate the impact of BALs based on the work of \cite{Ennesser2022}, which is to mask the locations of suspected BAL features in the \lya forest and exclude those pixels, although include the remaining pathlength. This is in contrast to most previous work, which removed the BAL quasars entirely \cite[e.g.][]{Bautista2017,dMdB2020}. The rationale behind the methodology proposed in \cite{Ennesser2022} is that BAL features are associated with concentrations of gas that have some range of outflow velocities, velocities that can range up to $\sim 0.1c$ from the systemic redshift of the quasar, and can have broad widths of many hundreds to thousands of $\mathrm{km\,s^{-1}}$. The velocity range of the absorbing material is relatively straightforward to measure in the vicinity of the \ion{C}{IV} emission line at $1549$\,\AA, and a conservative approach is to simply assume that some absorption is present in the same velocity range relative to other emission lines.

The analysis presented in \cite{Ennesser2022} quantified the gains from BAL masking with respect to the uncertainties in the correlation function, and showed that masking rather than complete elimination of the BALs results in a decrease in the uncertainties in the correlation function proportional to the fraction of BALs. Furthermore, the BALs introduce no systematic difference in the shape of the correlation function when they are masked. This paper extends the work of \cite{Ennesser2022} with a systematic analysis of the impact of masking on BAO measurements. In section~\ref{sec:data} we briefly describe the DESI observations and synthetic datasets or mocks that we analyze in this study. The fidelity of the mock spectra is important, as BALs are one of the astrophysical `contaminants' that make the mocks realistic. In section~\ref{sec:bals} we describe the main parameters of BALs, the algorithm that identifies them, the templates that we use to add BALs to mock data, and finally the completeness of the identification algorithm. We next evaluate how BAL features impact quasar redshift errors in section~\ref{sec:redshifts}. Redshift errors are potentially important for the cross-correlation measurement, as well as the quasar auto-correlation function. We present our main results in section~\ref{sec:results}, where we investigate the the continuum fits, correlation functions, and the BAO measurements for a range of BAL masking strategies. This includes an evaluation of how BAL quasar redshift errors affect the BAO measurements. We conclude in section~\ref{sec:summary} with a brief summary of our main results. 

\section{Data} \label{sec:data}

The first year data assembly of the DESI survey includes spectroscopic measurements of approximately 13 million galaxies, 1.5 million quasars, and approximately 4 million stars \cite{DESI2024.I.DR1} that form the basis for the DESI DR1 science results. A series of key papers present the large-scale structure catalogs \cite{DESI2024.II.KP3}, cosmological measurements at a range of redshifts \cite{DESI2024.III.KP4,DESI2024.V.KP5,DESI2024.IV.KP6}, and the cosmological implications \cite{DESI2024.VI.KP7A, DESI2024.VII.KP7B, DESI2024.VIII.KP7C}. In the first subsection, we briefly describe the DESI experiment that has enabled these results. This includes a description of the quasar catalog. We then provide a brief description of the mock datasets that play a critical role in the validation of the analysis methodology. This includes how BALs are added to these data. 

\subsection{DESI} \label{sec:desi}

The DESI experiment obtained more spectra in its first year of operations than all previous experiments. This unprecedented survey speed is due to a significant advances in instrumentation, superb calibration stability, and substantial software development. The DESI instrument is a highly multiplexed fiber spectrograph with 5000 fiber positioner robots \cite{FocalPlane.Silber.2023} located at the prime focus of the 4-m Mayall telescope at the Kitt Peak National Observatory. The 5000 fibers are located behind a six-element corrector system, including a two-element atmospheric dispersion corrector, with a $3^\circ$ diameter field of view \cite{Corrector.Miller.2023}. The fiber system \cite{Fibers.Poppett.2024} connects the focal plane system to ten, bench-mounted spectrographs that are maintained in a climate-controlled enclosure that provides excellent stability. Each spectrograph has three wavelength channels that together record the light from $360-980$\,nm at a spectral resolution that ranges from $2000-5000$. Further details of the instrument, including science and technical requirements, are described in \cite{DESI2022.KP1.Instr}. 

Numerous software tools and packages support the scientific and technical operations of the DESI survey. The DESI targets are based on the imaging dataset from the Legacy Surveys \cite{LS.Overview.Dey.2019}, and the target selection pipeline is described by \cite{TS.Pipeline.Myers.2023}. The spectroscopic pipeline is described in detail in \cite{Spectro.Pipeline.Guy.2023}. Some highlights of this pipeline include precise spectrophotometric calibration, noise estimates, sky substraction, and that fully processed data from each night are typically available to the collaboration the next morning. DESI survey operations plays a critical role in the very high efficiency of the experiment, including planning for each night of observations, automatic selection of fields during each night, and quality assurance the following morning. Survey operations are described in detail in \cite{SurveyOps.Schlafly.2023}. 

The quasar catalogs for the DESI DR1 analysis are largely comprised of quasar targets, although for \lya measurements we also include serendipitous discoveries of high-redshift quasars that were in other target classes, most notably emission line galaxies. The preliminary quasar target selection is described in \cite{QSOPrelim.Yeche.2020}. Prior to the start of the main survey, DESI had an approximately six month Survey Validation period \cite{DESI2023a.KP1.SV} to validate the selection of quasars \cite{Chaussidon2023} and other target classes, although also to optimize the instrumentation and operations prior to the start of the survey. The quasar validation process included a substantial visual inspection campaign described in \cite{VIQSO.Alexander.2023}. These results ultimately led to the use of three tools to identify quasars: the \texttt{Redrock} software that fits spectral templates and measures redshifts \cite{Redrock.Bailey.2024}, an \ion{Mg}{II} afterburner that searches for broad \ion{Mg}{II} emission in quasar targets that \texttt{Redrock} identifies as galaxies, and QuasarNet, a convolutional neural network classifier \cite{Busca2018}. Over the past year, we have improved the quasar templates used by \texttt{Redrock} to obtain more reliable classifications and redshifts \cite{Brodzeller2023}, and have improved the modeling of the \lya mean transmission \cite{KP6s4-Bault}. This work uses the same redshift catalog as the \lya DR1 BAO analysis. That catalog, including the BAL parameters, and the spectra will be publicly released with DESI DR1 \cite{DESI2024.I.DR1}. 

\subsection{Mocks} \label{sec:mocks}

The synthetic datasets that we use to study the impact of BALs were generated for the DESI DR1 data set. The construction of these mocks is very similar to the mocks that \cite{Herrera2024} produced for EDR, and that work describes the mock development in detail. The DR1 mocks have a few updates relative to those generated for EDR, which are described in the companion paper by \cite{KP6s6-Cuceu}. We therefore only provide a short summary of the generation of the mocks in this paper, and refer to those other works for more information. Section~\ref{sec:bals} has a description of how BALs are added to the mocks. 

The mocks are created in two stages. The first stage is the creation of the transmitted flux skewers for the sight lines to each quasar. This step uses a Gaussian random field to simulate the matter distribution, and the the quasar positions are from the log-normal transformation \cite[e.g.][]{coles91}. The second stage combines those transmission skewers with mock quasar spectra that are representative of the distribution of quasars in DESI. This includes a range of quasar spectral energy distributions and magnitudes, as well as the addition of noise and other astrophysical effects that make the spectra more realistic. The noise is added based on a model for instrument that includes the throughput and detector properties and the astrophysical effects include metal absorption and BALs \cite[see][for details]{Herrera2024,KP6s6-Cuceu}. The DESI \lya DR1 analysis uses two types of mocks referred to as \texttt{Ly$\alpha$CoLoRe} \cite{Farr2020_LyaCoLoRe} and \texttt{Saclay} \cite{Etourneau2023} mocks. In this paper we only consider the \texttt{Ly$\alpha$CoLoRe} mocks because BALs are added in exactly the same way to both types of mocks. The differences between these two types of mocks include factors such as how the quasar distribution and velocity field are modeled. 

The density and velocity distributions for the transmitted flux skewers use the \texttt{CoLoRe} package \cite{Ramirez2022} to generate Gaussian random fields and the Newtonian potential of this field to determine the velocity field for the skewers. We then convert skewers through this density and radial velocity field into skewers of transmitted flux with the \texttt{Ly$\alpha$CoLoRe} \cite{Farr2020_LyaCoLoRe} package. This package adds additional, small scale information based on a one-dimensional Gaussian and computes the optical depth with the fluctuating Gunn-Peterson approximation \cite{Bi1997,Croft1998}, as well as adds redshift space distortions based on the radial velocity field. 

These transmitted flux skewers are added to a quasar population that matches the magnitude, redshift, sky distribution, and density of objects on the sky of the DESI DR1 dataset. One update in DR1 relative to the EDR mocks is a change in the way the mock distribution samples inhomogeneities in the observational data, which is described in detail in \cite{KP6s6-Cuceu}. We use the \texttt{quickquasars}\footnote{https://github.com/desihub/desisim/blob/main/py/desisim/scripts/quickquasars.py} script from the \texttt{desisim}\footnote{https://github.com/desihub/desisim} package to generate a synthetic quasar spectrum for each quasar, add astrophysical features to make the spectra more realistic, and then add the appropriate level of noise based on the magnitude of each quasar and the number of observations of that quasar available for DR1. The astrophysical features include Damped Lyman $\alpha$ systems (DLAs), BALs, and absorption from metals in the IGM, specifically the \ion{Si}{II} $\lambda 1190$, \ion{Si}{II} $\lambda 1193$, \ion{Si}{III} $\lambda 1207$, and \ion{Si}{II} $\lambda1260$ lines that are most important absorption features in the forest region. The DLAs and metals are added based on the same density field used to generate the transmitted flux skewers, although the relative strengths of the metal lines are tuned following a procedure described in \cite{KP6s6-Cuceu} to match the observational data. The BALs are randomly applied to 16\%. This percentage is based on measurements from SDSS and early DESI data \cite{Gibson2009,Filbert2023}. 

\section{Broad Absorption Line Quasars in DESI} \label{sec:bals}

\begin{figure}[htbp]
\centering
\includegraphics[width=1.0\textwidth]{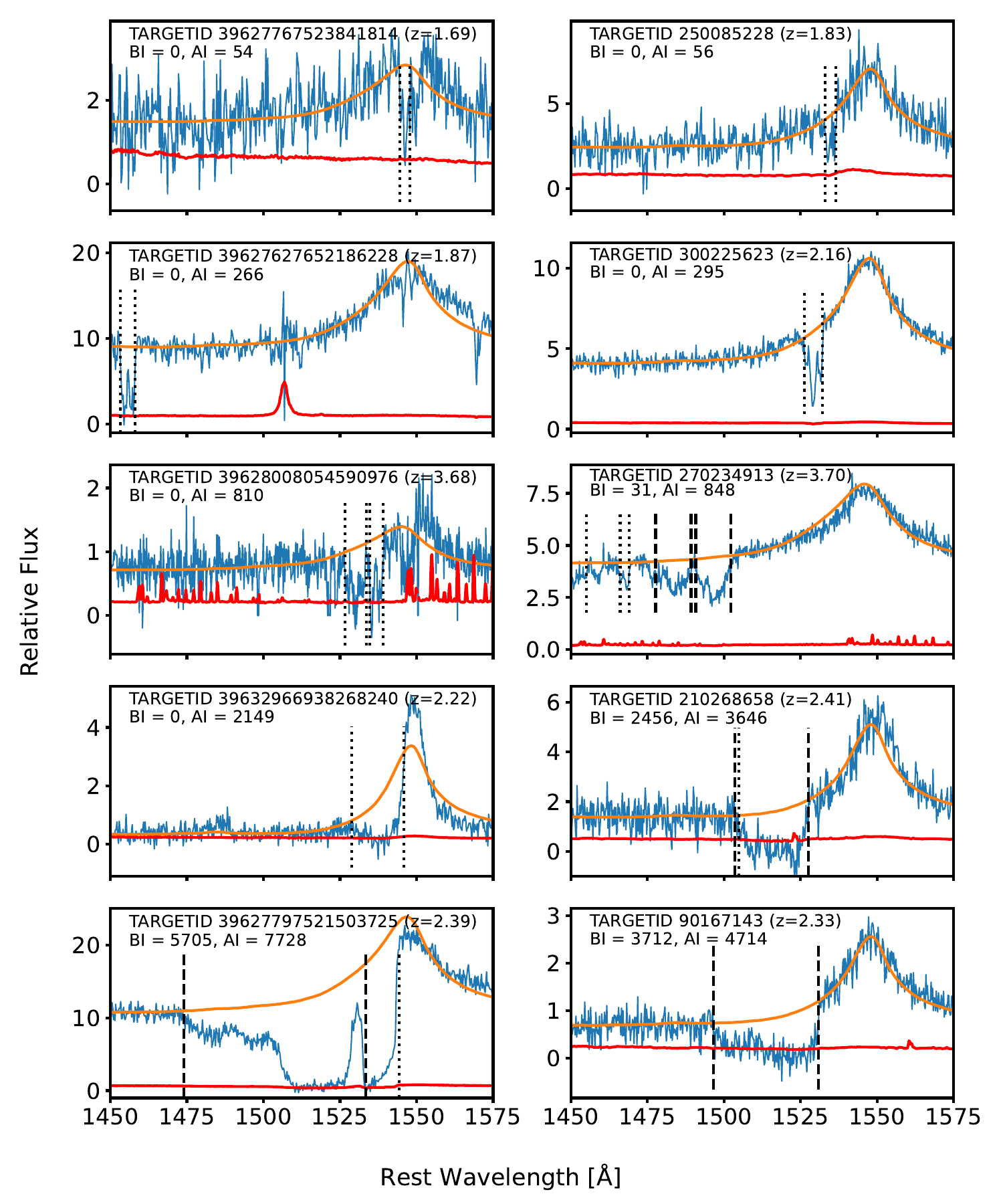}
\caption{Selection of BALs in DESI DR1 data ({\it left column}) and mocks {\it right column}) with a range of AI values. The spectra ({\it blue lines}) are centered on the wavelength range around the \ion{C}{IV} region that is used to identify BALs. Also shown are the template fit used to identify BAL troughs ({\it orange line}) and the error in the flux ({\it red line}). The vertical lines mark the limits of the troughs that meet the AI criterion ({\it dotted}) and BI criterion ({\it dashed}). Each row shows a randomly selected pair of real and mock BAL quasars with similar AI values (or BI values, for the last row). The top four rows represent the four quartiles of the AI distribution of the data. The bottom row shows a real and a mock BAL with significant BI values. The TARGETID, redshift, AI, and BI values are listed in each panel (TARGETID is a unique identifier used by DESI for each target). \label{fig:spectra} }
\end{figure}

We use templates to add BALs to the mocks, rather than a series of parametric functions, due to the inherent complexity and diversity of BAL features. Figure~\ref{fig:spectra} shows several BALs from DESI DR1 observations and from mocks. The BALs added to the DR1 mocks are based on approximately 1500 templates calculated from BAL quasars by \cite{Niu2020}. These BAL quasars are a subset of those identified in SDSS DR14 data by \cite{Guo2019}. In the first subsection, we describe common parameterizations for BALs and the development of the templates. We then briefly summarize the BAL identification algorithm and the data record for each BAL quasar. This identification does not identify all BALs, and also has some false positives, and we characterize the completeness and purity of the algorithm with a study of the mocks. Finally, we use the much larger BAL dataset from DESI DR1 to examine the fidelity of the observed BALs to the mock datasets. 

\subsection{Parameters} \label{sec:params}

The broad absorption features characteristic of BAL quasars appear as one or more troughs that almost always appear on the blue side of the broad emission lines, especially higher ionization lines such as \ion{C}{IV}. The original parameter used to characterize and compare BALs is the Balnicity Index (BI) proposed by \cite{Weymann1991}. The equation for BI is: 
\begin{equation}\label{eq:BI}
    BI = - \int_{25000}^{3000} \left[ 1- \frac{f(v)}{0.9} \right] C(v) dv\rm{.}
\end{equation}
The variable $v$ is the velocity relative to the nominal central wavelength of the emission feature, $f(v)$ is the observed flux distribution of the quasar divided by a model of the quasar if the BAL features were not present, and $C(v)$ is a function that is zero unless the term $(1 - \frac{f(v)}{0.9})$ is greater than zero for more than $2000$\,km\,s$^{-1}$, in which case it is set equal to one. BI is consequently similar to an equivalent width, where the difference is the requirement that the trough extend for at least $2000$\,km\,s$^{-1}$ before the start of the integral over the absorption. The rationale for this choice was to ensure that the absorption was much broader than could be explained by galaxy kinematics, and the integration limits eliminate absorption that could be due to the host galaxy. 

Studies of progressively larger numbers of BALs with SDSS showed that many quasars have broad absorption that extends closer to the quasar rest frame than $3000$\,km\,s$^{-1}$, and that have widths less than $2000$\,km\,s$^{-1}$, yet are clearly still broader than expected from normal motions within galaxies. This prompted \cite{Hall2002} to propose the Absorption Index (AI). The equation for AI is: 
\begin{equation}\label{eq:AI}
    AI = - \int_{25000}^{0} \left[ 1-\frac{f(v)}{0.9} \right] C(v) dv
\end{equation}
The two main differences from BI are that the integration extends to the systematic redshift and the function $C(v)$ is set to one after the trough extends for only $450$\,km\,s$^{-1}$. There is no other distinction, for example related to the depth of the absorption, as both indices require that more than 10\% of the quasar flux is absorbed.  

The AI criterion captures many more BALs than the BI criterion, yet is the appropriate criterion to use as BALs identified based on the AI absorption feature can have similar total depth as the BI features, even if they do not extend over as large a range in wavelength. And only the AI definition captures BAL features that impinge on the strong emission features like \ion{C}{IV} that are an important part of quasar redshift measurements. 

While the AI and BI parameters are broadly useful to capture the relative amount of total absorption due to BALs, these single parameters do not adequately capture the full diversity of BALs. In addition to the total absorption, other quantities that vary between BALs are the blueshifts of the minimum and maximum velocity of each trough, the variation in absorption with wavelength within each trough, the number of troughs per quasar, and the relative strength of the troughs associated with different emission features. These variations defy simple parameterization, and we consequently developed a set of 1500 empirical templates to add BALs to mock datasets starting with the work of \cite{Niu2020}. 

The templates built in that work, and later refined as described by \cite{Filbert2023}, started with a sample of about 1500 very high signal-to-noise BAL quasar spectra identified by \cite{Guo2019} in SDSS DR14. These templates appeared broadly representative of real BALs, as for example the AI and BI distributions of the parent sample of BALs were consistent with the distributions in the full DR14 BAL population. Those previous works then fit each BAL with a continuum model after masking the BAL features, normalized the spectra after dividing by the continuum model, and set the template equal to one outside the BAL regions so that the templates did not add unnecessary noise to mock quasars. This produced a model for the normalized absorption features of the BALs associated with \ion{C}{IV}. 

The templates for the BAL absorption are calculated based on observed absorption in the \ion{C}{IV} region and are applied to other potential emission features with the assumption that the absorption vs.\ velocity profile of the BALs are independent of element and ionization state. In reality this is not true in detail, as BAL absorption is a complex function of the ionization state and velocity of the absorbing gas, as shown with significant modeling efforts to understand the physical conditions in well-studied, high SNR BAL spectra \cite{Leighly2019}. Since the cosmological analysis masks the BAL absorption regardless of its structure, it only matters that the velocity distribution is approximately the same for different ions. 

We use the BAL stacking study of \cite{Masribas2019} to determine the emission lines that are observed to have BAL features. The BAL templates used for the DR1 analysis add BAL absorption associated with \ion{S}{IV}, \ion{N}{V}, \lya, \ion{C}{III*}, \ion{P}{V}, \ion{S}{IV}, and \lyb. Based on the BAL stacks of \cite{Masribas2019}, the \ion{C}{III*}, \ion{P}{V}, \ion{S}{IV}, and \lyb lines are substantially weaker, and the absorption in these lines was set to 10\% of the absorption present in \ion{C}{IV}. This is relevant for our experiments on mocks with different masking choices in Section~\ref{sec:results}.

\subsection{Identification Algorithm} \label{sec:algorithm}

The BAL identification algorithm we use for DESI DR1 and mocks is nearly identical to the one used by \cite{Filbert2023} for EDR, so we only briefly summarize it here. The algorithm is based on fitting a series of templates to every spectrum around the rest-frame \ion{C}{IV} emission feature and searching for absorption relative to this model fit. We set the blue limit of this fit to 25,000\,km\,s$^{-1}$ blueshift (about rest frame 1420\,\AA) relative to the \ion{C}{IV} line. For most quasars the spectral range extends to 2400\,\AA\ and we set a minimum of at least to 1633\,\AA\ for the highest redshifts. This sets a redshift range of $1.57 < z < 5$ based on the observed wavelength range of the DESI data. We shift the spectrum to the rest frame based on the redshift calculated by \texttt{Redrock}. This redshift uses a prior from one of the other tools described in \ref{sec:desi} if Redrock did not initially identify the object as a quasar. 

We use the best spectral fit to search for BALs associated with \ion{C}{IV} based on the AI and BI criteria described in section~\ref{sec:params}, namely with the ratio of the observed flux to the best spectral fit. Should a BAL feature be present, we iteratively mask the velocity range of the trough and refit the templates for either ten iterations or convergence. The iteration process improves the quality of the continuum fit outside of the trough region(s). If there is sufficient spectral coverage, we also perform a separate search in the vicinity of the \ion{Si}{IV} emission feature. We then record the AI and BI values, velocity range of each trough, and other parameters as described in \cite{Filbert2023}. 

We use the public \texttt{baltools}\footnote{https://github.com/paulmartini/baltools} software package to identify and measure the BALs in DESI. This code was originally developed by \cite{Guo2019} to measure the parameters of BALs that were identified via their convolutional neural network, which they applied  to SDSS DR14 data. The code was later applied to identify BALs for the DR16 quasar catalog \cite{Lyke2020}, although that worked dropped the CNN component for classification and instead relied on just the measurement of AI and BI to identify BALs with the algorithm described in this subsection. This approach was also adopted by \cite{Filbert2023} for the DESI EDR BAL catalog, and we continue that approach here. 

The most significant change in the \texttt{baltools} package since the EDR catalog is the change from PCA templates calculated by \cite{Guo2019} to new quasar templates developed by \cite{Brodzeller2022}.  These new templates perform better than the old PCA templates, including the fit to the \ion{C}{4} region, especially for high SNR spectra, and lead to fewer false positives in high SNR data. For example, based on visual inspection the number of false positives at high SNR (SNR$>5$) decreased from about 10\% to consistent with zero. 

\begin{figure}[htbp]
\centering
\includegraphics[width=1.0\textwidth]{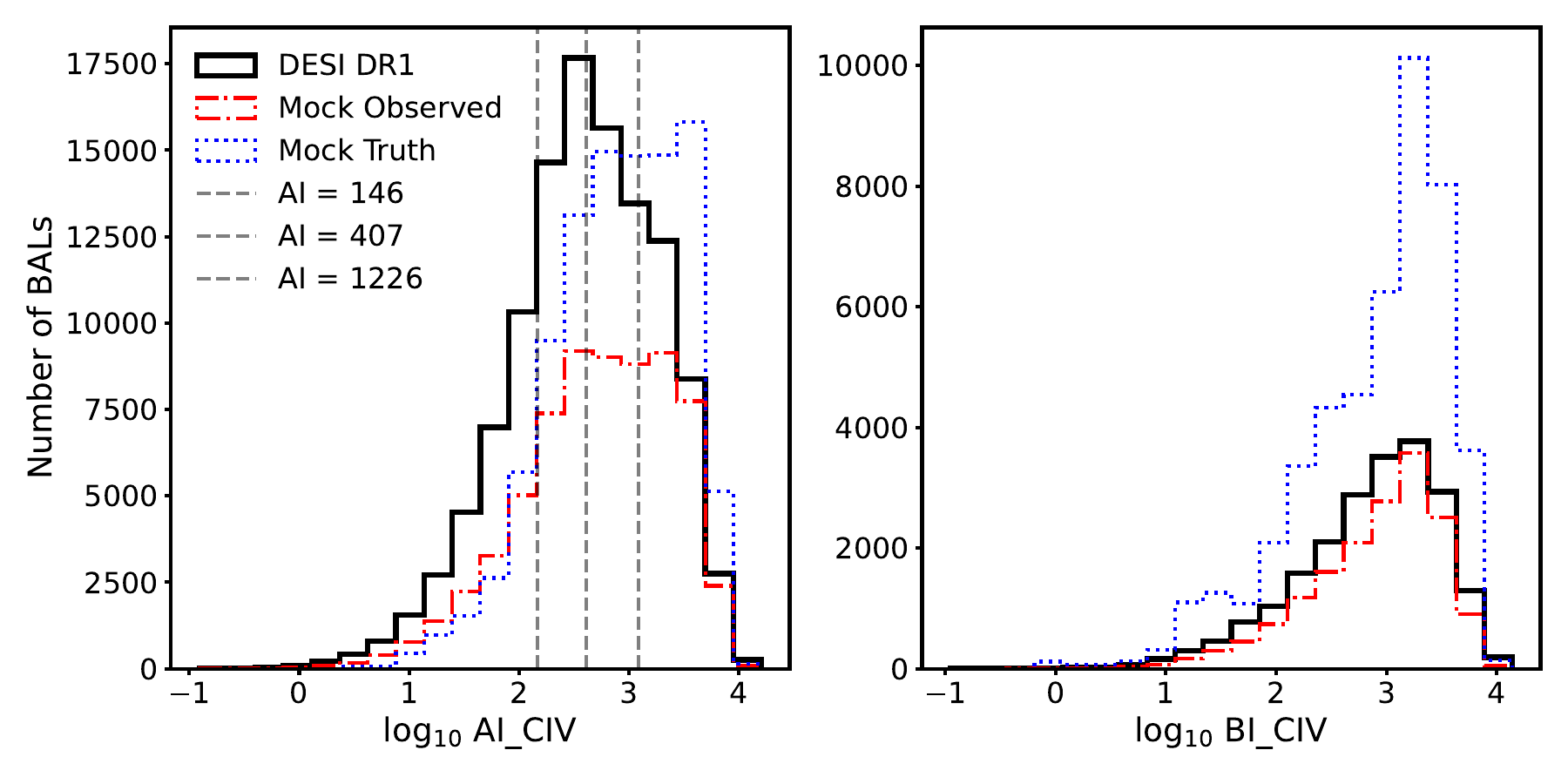}
\caption{Distribution of AI and BI values for the DESI DR1 data ({\it black, solid histogram}) and DR1 mock datasets. The histograms for the mock datasets show both the distribution returned or observed by the BAL identification algorithm ({\it red, dashed}) and the true distribution ({\it blue, dotted}). Three vertical lines mark the AI values that separate the AI distribution into four quartiles. These are located at $AI = 146, 407$, and $1226$. We discuss the discrepancy between the data and mocks in \S\ref{sec:balmock}. \label{fig:aibidist} }
\end{figure}

We used the \texttt{baltools} package to search for BALs associated with every DESI DR1 quasar in the redshift range $1.57 < z < 5$. The BAL fraction is 16.7\% in the redshift range $1.8 < z < 3.8$ used for the \lya analysis based on the criterion $AI > 0$ and 1.3\% based on the criterion $BI > 0$. These percentages are similar to those measured by \cite{Filbert2023} based on the first two months of the main survey, although the AI percentage is slightly higher and the BI percentage is slightly lower. The AI and BI distributions of the BALs are shown in figure~\ref{fig:aibidist}, which also shows divisions of the AI distribution into four quartiles that are separated by $AI = 146$, $AI = 407$, and $AI = 1226$. These values are similar although somewhat smaller than the values for the SDSS distribution \cite[250, 839, and 2221, respectively, see][]{Ennesser2022}, which is likely due to the different SNR and perhaps other features of the data. 

\begin{table}[htbp]
    \centering
\begin{tabular}{l|r|r|r|r}
    \hline
     Option & DESI DR1 Number & DESI DR1 \% & Mock Observed \% & Mock Truth \% \\
     \hline
$AI > 0$ (baseline) & 112822 & 16.7 & 10.6 & 15.8 \\
$AI > 146$ & 84922 & 12.6 & 8.5 & 14.0 \\
$AI > 407$ & 57086 & 8.4 & 6.2 & 10.9 \\
$AI > 1226$ & 28575 & 4.2 & 3.6 & 6.5 \\
$BI > 0$ & 8650 & 1.3 & 1.2 & 3.6 \\
$AI > 0$ (masked z) & 112554 & 16.6 &  N/A &  N/A \\
     \hline
\end{tabular}
    \caption{Fraction of quasars with $1.8 < z < 3.8$ that exhibit BAL features based on the masking option listed in column one. For each option, we list the total number of BALs in the DESI DR1 data in this redshift range, the fraction of DESI DR1 quasars that are BALs with this option, the fraction observed in the mock catalog with the selection algorithm, and the true fraction in the mock catalog. The values of 146, 407, and 1226 separate the four quantiles of AI in the DESI DR1 data, as shown in Figure~\ref{fig:aibidist}. The last row has the BAL fraction in DESI DR1 data when we run a second iteration of \texttt{Redrock} on the BALs after masking their absorption troughs. There is a very small ($0.1$\%) decrease in the BAL fraction. }
    \label{tab:balfrac}
\end{table}

We note that since all of these BALs were identified in the vicinity of the \ion{C}{IV} (or \ion{Si}{IV}) emission feature, they all appear to be high-ionization BALs or HiBALs. While we cannot rule out the presence of BAL absorption associated with lower ionization features such as \ion{Mg}{II} (known as LoBALs) or even significant iron absorption \citep[FeLoBALs, and see][for some examples of these two later types]{Filbert2023}, HiBALs are by far the most common. For example, \cite{Trump2006} found that 26\% of all quasars are HiBALs, 1.3\% are LoBALs, and 0.3\% are FeLoBALs. Furthermore, LoBALs and FeLoBALs also typically exhibit \ion{C}{IV} absorption, although those BALs can also exhibit such extreme absorption that the automated redshift fitting algorithms do not work well. 

\subsection{Completeness and Purity} \label{sec:comp}

The completeness and the purity of the \texttt{baltools} detection algorithm is an important measure of the fraction of BALs that are missed, as well as the fraction of quasars that are incorrectly classified as BALs. The work by \cite{Filbert2023} measured the completeness of the algorithm by analyzing mock spectra in the same manner as the observations. That work found that the completeness was 68\% for the mock catalog, and that most of the missed BALs were in data with $SNR < 2$ per pixel. This is lower completeness than estimated with the CNN classifier by \cite{Guo2019} and earlier work by \cite{Busca2013} with SDSS data, although both of those earlier works relied on human classified BALs for their truth catalogs, and thus the BAL samples were biased towards BALs that are obvious to a human. 

\begin{figure}[htbp]
\centering
\includegraphics[width=1.0\textwidth]{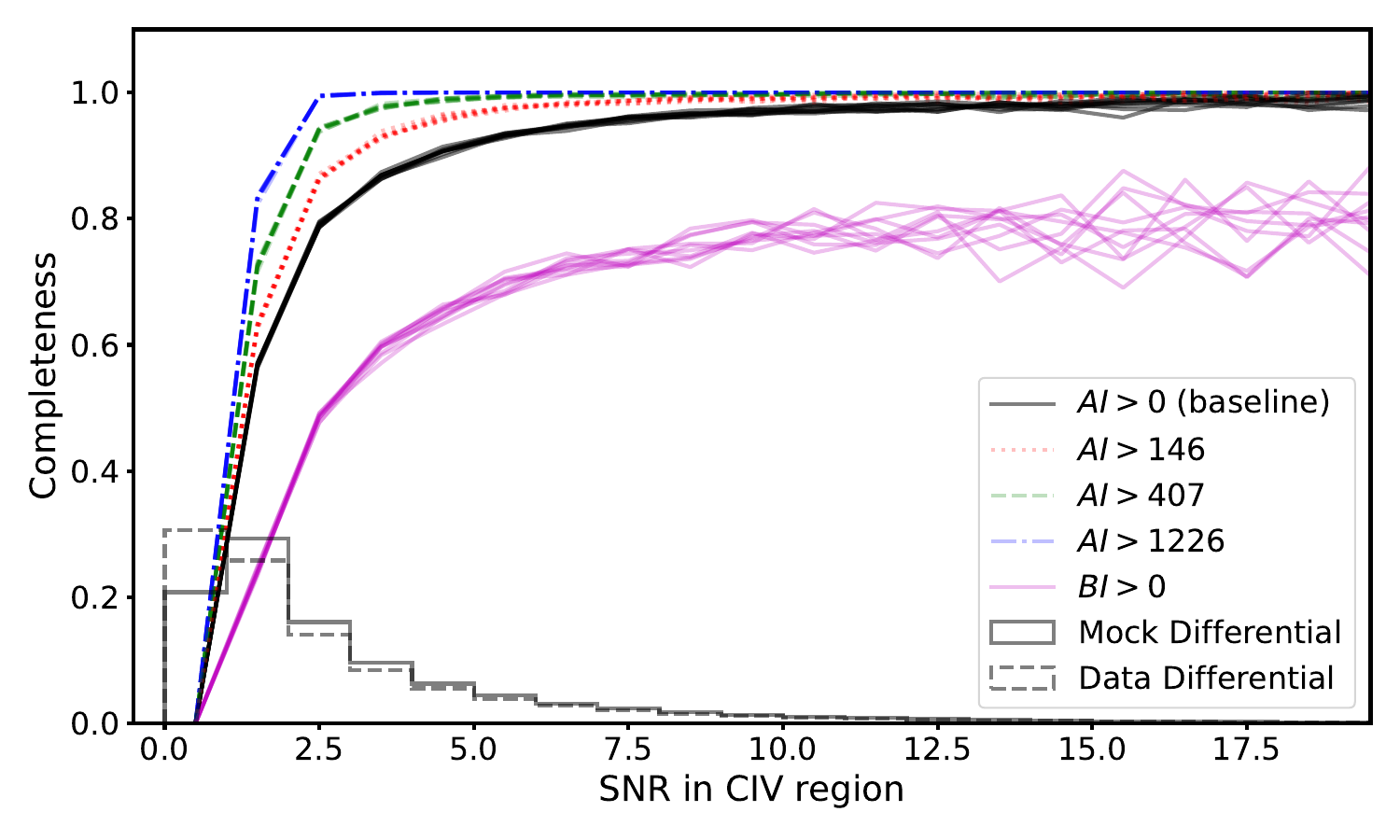}
\caption{Completeness and purity of the BAL identification algorithm as a function of SNR in the \ion{C}{IV} region based on ten DESI DR1 mock datasets. The completeness is a strong function of SNR, ranging from 18\% in the first SNR bin to 100\% for the highest SNR spectra. The completeness is also a strong function of AI, in that the completeness is higher for larger AI values. The completeness for features that meet the BI only reaches 80\%, which is because the troughs do not meet this criterion, rather than the quasars are not classified as BALs. The distribution of the data ({\it dashed histogram}) and the mocks ({\it solid histogram}) is also shown, and indicates most quasars are low SNR. Numerical values for several of these quantities are listed in Table~\ref{tab:cp}. }  \label{fig:cp}
\end{figure}

We revisit the completeness (and purity) analysis from \cite{Filbert2023} with a much larger set of mocks that are designed to match the DESI DR1 dataset. Figure~\ref{fig:cp} shows the average completeness as a function of SNR per pixel for ten DESI DR1 mocks and the numerical values are listed in the Appendix in table~\ref{tab:cp}. We calculated this quantity by running the BAL identification algorithm on the mock spectra and comparing the observed BAL catalog for each mock to the truth BAL catalogs. All of the mocks clearly confirm that the completeness of the algorithm is a strong function of SNR with little variation between mocks. This also confirms the point made by \cite{Ennesser2022} that BALs are preferentially associated with higher SNR spectra, which compounds the advantage of keeping these spectra for \lya analysis and just masking the locations of their absorption troughs. The completeness is 60\% for the mocks, which is somewhat higher than the 42\% estimated by \cite{Filbert2023} for the first two months of the main survey. This is likely because the typical SNR of the DR1 mocks is somewhat higher, somewhat over 40\% of the \lya quasars in the DR1 sample have been observed more than once. The cumulative completeness of the data will be lower than measured for the mocks, as the SNR of the mocks is somewhat higher than the data. The cumulative completeness is predicted to be about 53\% if we weight the differential completeness as a function of SNR by differential distribution of the data as a function of SNR (columns 2 and 4 of table~\ref{tab:cp}).

The figure also shows that the completeness is a function of AI value, in the sense that BALs with larger AI values are easier to spot in lower SNR data than BALs with lower AI values. This result suggests that incompleteness may not have an impact on the measurement of the correlation function, nor on the BAO parameters, as the missed BALs have relatively small amounts of absorption and are in the lowest SNR data that have the smallest weights in the correlation functions. 

We have used the same mock BAL catalogs to measure that the purity is approximately 90\% (see Table~\ref{tab:cp}). The purity actually drops as SNR increases, which we did not see in our previous analyses of mocks without redshift errors. Based on our visual inspection of some of these cases, it appears that the continuum fits are not as good as in the absence of redshift errors. This is because the BAL detection algorithm does not attempt to refit the redshift, and as a result there is some contamination. Nevertheless, this is a very small effect as fewer than 10\% of the quasars are in this regime with lower purity. It is also unlikely to be an issue in the data, as the redshift measurements in the data will be tied to strong emission features like \ion{C}{IV}. Lastly, the excellent performance at low SNR indicates that the algorithm does not tend to erroneously classify noise as BAL absorption. We consequently expect that there is a correspondingly small fraction of path length that is being masked unnecessarily. One caveat is that the mocks may not include all of the other astrophysical features that may mimic the appearance of BALs, such as metal absorption features in the ISM or more intrinsic quasar diversity. Therefore this analysis may have somewhat overestimated the purity compared to real data. 

\subsection{BAL Mock Fidelity} \label{sec:balmock}

The BAL templates that we employ for the DESI DR1 analysis were developed from quasars in the SDSS DR14 quasar catalog \cite{Paris2018}, and collectively are a good match to the AI and BI distributions of that dataset. The total number of quasars cataloged by DESI in the DR1 sample is approximately three times larger than the SDSS DR14 catalog. In addition, the SNR per spectrum and the spectral resolution are different between the two surveys. Preliminary studies \cite{Filbert2023,Herrera2024} have shown that the AI distribution of the DESI EDR data was different from SDSS DR14 and from the mocks, while the BI distributions were nearly identical. Here we use the larger dataset from DESI to evaluate further if the BAL templates in our mocks provide a realistic description of the BALs observed by DESI. 

Our first point of comparison is the distribution of AI and BI values in the data relative to the mocks. Figure~\ref{fig:aibidist} shows the distributions of both quantities for DESI DR1, along with histograms of the true and observed distributions from the mocks. The observed AI distribution from the data and the mocks are not in good agreement, in that there are many fewer BALs recovered by the algorithm for the mock dataset than DESI observations. In contrast, the true distribution in the mocks is a much better match to the DESI data, although the median AI value is somewhat larger in the mocks than in the data. This difference may in part be due to the somewhat lower SNR per spectrum of the DESI data relative to SDSS. This will especially be the case for quasars at $z < 2.1$, which DESI only observes once. While DESI ultimately aims to obtain multiple observations of the \lya quasar sample at $z>2.1$, most of these quasars were only observed a single time during the first year of observations. 

The agreement between the DESI BI distribution and the mocks is excellent. While the true distribution of BI values is higher than in the data, the distribution of the mock BALs recovered by the algorithm is extremely similar to the DESI observations. The discrepancy between the AI and BI distributions is consequently somewhat surprising, as there are not separate templates for troughs that meet the AI and BI criteria. 

\begin{figure}[htbp]
\centering
\includegraphics[width=1.0\textwidth]{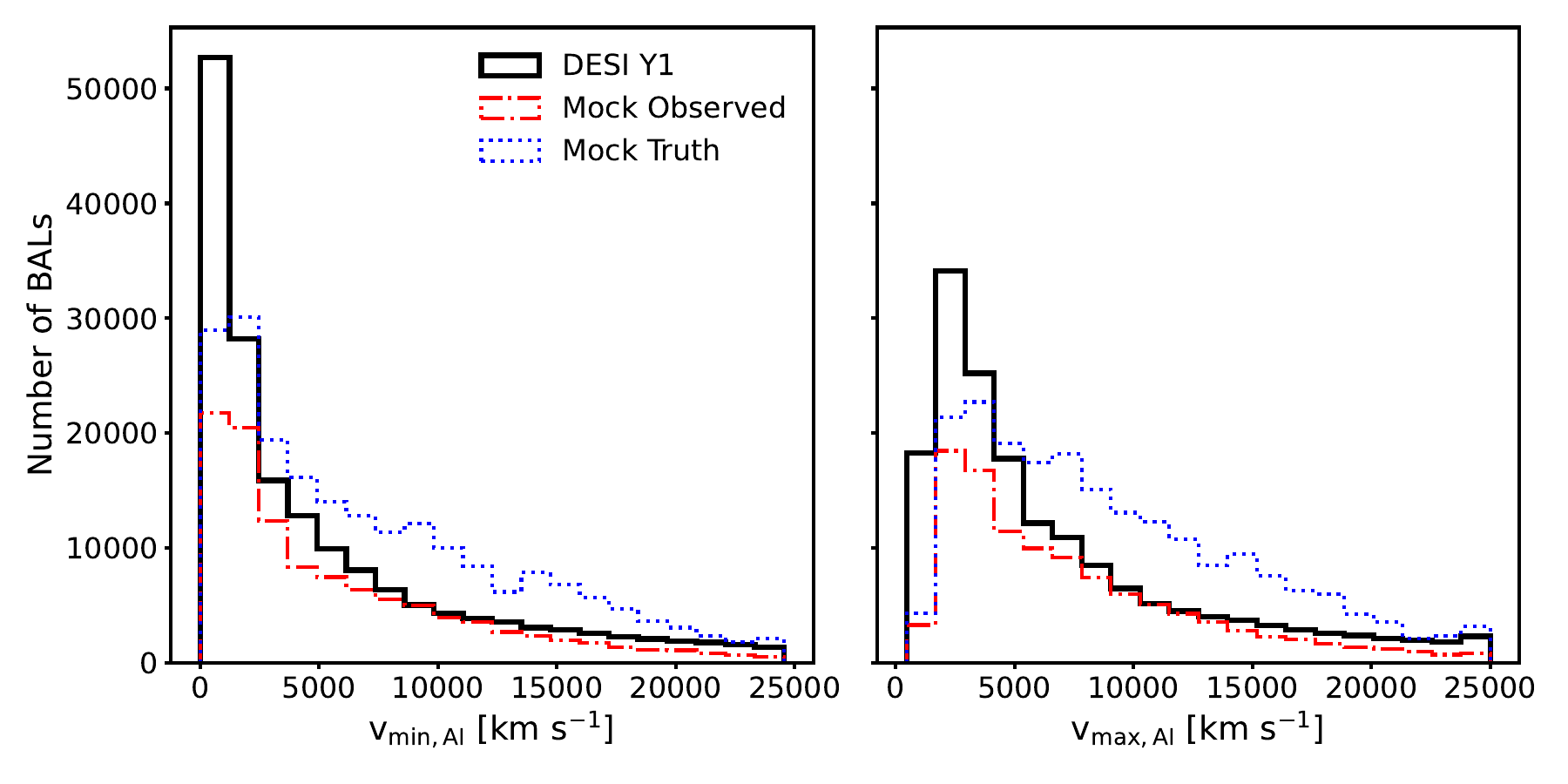}
\caption{Distribution of the trough velocities for AI troughs of the DESI DR1 data ({\it solid histogram}), the mock catalog produced by the BAL identication algorithm ({\it dashed}), and the truth catalog for the DR1 mock ({\it dotted}).  \label{fig:vdist}}
\end{figure}

Figure~\ref{fig:vdist} compares the velocity distributions of the troughs between the mocks and the data. These quantities are important because they establish the velocity range for the pixels mask in our analysis. The two panels show the minimum and maximum blueshift velocity of each that meets the AI criterion (there are two troughs per BAL on average). These distributions for the DESI data and the observed mocks are in very good agreement, and the mock truth is somewhat higher as expected based on the completeness of the algorithm. The exception is that there appear to be somewhat fewer BAL troughs that start at very low blueshifted velocities, both relative to the data and truth. This may indicate that the templates had a selection bias against BAL features that were part of the blue wing of the \ion{C}{IV} emission feature, which is plausible because it is difficult to model the blue wing of that line and the templates were derived with the PCA components developed by \cite{Guo2019}, while the DESI observations used new components derived by \cite{Brodzeller2023}. 

\section{Impact on Redshift Errors} \label{sec:redshifts}

\begin{figure}[htbp]
\centering
\includegraphics[width=1.0\textwidth]{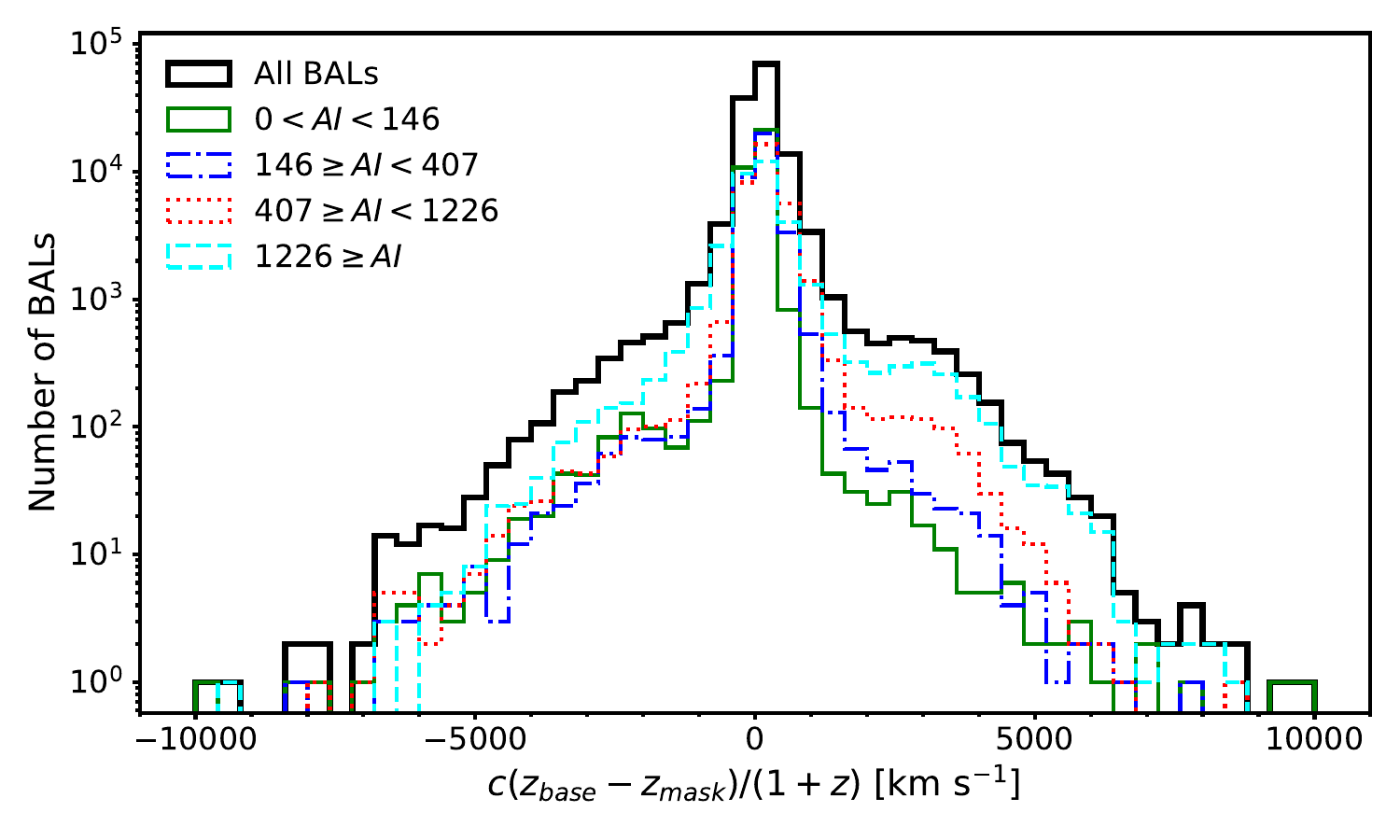}
\caption{Distribution of velocity shifts for DESI DR1 BALs after masking the BAL features ({\it solid black histogram}). The shifts are more common and more shifted to larger values (lower redshifts) for BALs with larger AI values. The mean and median velocity shifts are listed in table~\ref{tab:maskz}. \label{fig:zdist} }
\end{figure}

\begin{table}[htbp]
    \centering
\begin{tabular}{l|r|r}
    \hline
     Criterion & Mean & Median \\
     \hline
	All BALs & 119.2 & 22.2 \\
	$0 < AI < 146$ & 9.7 & 3.0 \\
	$146 \leq AI < 407$ & 179.3 & 29.0 \\
	$407 \leq AI < 1226$ & 278.8 & 120.0 \\
	$1226 \leq AI$ & 271.9 & 52.0 \\
     \hline
\end{tabular}
    \caption{Redshift changes after rerunning \texttt{Redrock} with the BAL troughs masked. The changes are in $c(z_{base} - z_{mask})/(1+z)$ in units of km\,s$^{-1}$, as in figure~\ref{fig:vdist}.}
    \label{tab:maskz}
\end{table}

The histograms shown in Figure~\ref{fig:vdist} give a very good indication of how often the BAL features impinge upon the blue wing of the \ion{C}{IV} emission feature, which typically has a width of many thousands of km\,s$^{-1}$. This absorption can significantly change the shape of the line, and in the most extreme circumstances may completely absorb the entire blue wing. At the redshifts where DESI employs quasars to trace large scale structure, namely above about $z>1.5$, the \ion{C}{IV} line is one of the strongest spectral features within the observed wavelength range. Substantial asymmetric changes in the shape of this feature can consequently lead to redshift errors, including systematic offsets relative to the true redshift. 

The impact of BALs on redshift errors was thoroughly studied with mock DESI spectra by \cite{Garcia2023}. The redshift error or velocity shift is defined as: 
\begin{equation}
    c(z_{base} - z_{mask})/(1+z)
\end{equation}
\noindent
where $z_{base}$ is the measureded redshift without masking and $z_{mask}$ is the redshift measured after masking the BAL features. \cite{Garcia2023} showed that the redshifts derived by \texttt{Redrock} for mock spectra with BALs had offsets of $\Delta z \sim 2 \times 10^{-5}$ relative to the same mock quasar spectra without the BAL features. They then masked the locations of pixels that were potentially impacted by BALs, specifically \ion{C}{IV}, \ion{Si}{IV}, \ion{N}{V}, and \lya by setting the inverse variance of pixels to zero if they were at the same velocity relative to each feature as the BAL absorption associated with \ion{C}{IV}. The redshift offsets for the masked BALs were substantially smaller than for the unmasked BALs, where in both cases the redshift offsets were relative to the redshift if the BAL features were not present. Those authors found that masking the locations of the BAL features led to a substantial reduction in the redshift errors relative to the true value, as well as improvements in other areas such as the number of catastrophic redshift errors and spectral misclassifications. 

The masking strategy proposed by \cite{Garcia2023} was implemented by \cite{Filbert2023} as part of their study of BALs in the DESI EDR. While there is no true redshift in these cases, they measured the redshift difference before and after masking the BAL features and showed that the average velocity difference averaged over all BALs is $243$\,km\,s$^{-1}$, and that the size of the offset depends on AI value, such that the offset ranged from 56\,km\,s$^{-1}$ for the lowest quartile of AI value to 582.3\,km\,s$^{-1}$ for the highest quartile. In all cases the average velocity shifts are to larger values, which correspond to a decrease in the redshift, which is expected as BALs impact the blue wing of the emission features. Furthermore, 6.7\% of quasars had velocity shifts greater than $\Delta v > 1000$\,km\,s$^{-1}$, which is defined as a catastrophic redshift error for the quasars used to trace large scale structure \cite{DESI_SV}. 

We have repeated the masking analysis performed by \cite{Filbert2023} for the DESI DR1 data to measure the typical velocity differences between masking and not masking BALs, as well as to evaluate the impact of these errors on the BAO parameters. Figure~\ref{fig:zdist} shows the change in velocity for the entire BAL sample, as well as distributions for the four quartiles in AI values. The average and median offsets range from nearly zero for the lowest quartile to $150-200$ \,km\,s$^{-1}$ for the largest quartile. These offsets are substantially smaller than those found by \cite{Filbert2023}, which is likely because the DR1 quasar redshifts were measured with new quasar templates from \cite{Brodzeller2023}. That paper refit quasars from the \cite{Filbert2023} study with their new quasar templates and measured the cross-correlation function between the quasars with BALs and \lya absorbers to measure the average bias in the quasar redshifts. They found that the average shift was $\Delta r_{||} = -56 \pm 47$\,km\,s$^{-1}$ without masking with the new templates, as compared to $-177 \pm 63$\,km\,s$^{-1}$ without masking with the previous quasar templates, which is roughly consistent with the change between the average of $243$\,km\,s$^{-1}$ measured by \cite{Filbert2023} with EDR data and the previous quasar templates and the average of $119$\,km\,s$^{-1}$ we measure with the DR1 data and the new templates. 

The relative shift in the redshifts with the different templates and with BAL masking was also studied by \cite{KP6s4-Bault} with DESI EDR data. That study used the cross correlation between quasars and the \lya forest to measure redshift offsets associated with BALs before and after the BALs were masked by \cite{Filbert2023}. They found that after masking, the mean redshift offset of the BAL quasars agreed with non-BAL quasars within $0.35\sigma$. This result demonstrates that the redshifts after masking are more accurate and are not just different.

\section{Impact on Baryon Acoustic Oscillations} \label{sec:results}

Nearly 17\% of the DESI DR1 quasars with $1.8 < z < 3.8$ are observed to have BAL features associated with \ion{C}{IV} based on the AI criterion (see table~\ref{tab:balfrac}), and these BAL features add absorption in the forest region and impact redshift errors. In this section we evaluate the impact of BALs on the measurement of the location of the BAO peak with a range of alternative masking strategies. There are two motivations for this study. First, the BAL masking strategy we employ is quite conservative, in that we mask pixels in the velocity range of every BAL features identified in the stacking study of \cite{Masribas2019}. Most of these BAL features are expected to be quite weak, and may be negligible compared to the forest absorption, so the masking strategy may remove pathlength unnecessarily. We consequently fit the BAO peak with a range of alternative strategies that mask fewer BALs. These strategies are to only mask BALs above the AI values that divide each quartile in the data (see figure~\ref{fig:aibidist}), that is above $AI>146$, $AI>407$, and $AI>1226$. In each of these cases we mask BALs with AI values above this threshold and do not mask any features in BALs below this threshold. We also consider the case where we just mask BALs with $BI > 0$, as well as show the case $AI > 0$ (baseline) and $AI > 0$ (masked z), which has updated redshifts as described in Section~\ref{sec:redshifts}. The second motivation is that we know the BAL identification algorithm is incomplete based on our studies with mocks, and therefore there are unidentified BALs in our data that may compromise the analysis. We use the mock data to compared the results between catalogs based on the true BAL distribution and the `observed' distribution returned by the identification algorithm. 

There are three main steps in the \lya analysis that extracts BAO measurements from the spectra of quasars: 1) calculation of the variation of the absorption along the line of sight; 2) measurement of the correlation functions; 3) calculation of the BAO and other model parameters that best match the observed correlation functions. We expect BALs to impact each step because the masking process removes pixels, and unmasked (missed) BALs add contamination. In the first subsection, we briefly describe the continuum fitting process and quantify the impact of BALs. We then present measurements of the correlation functions for each masking option, as well as evaluate the analogous measurements with the mock data. Lastly, we fit the correlation functions with a cosmological model and evaluate the impact of different masking strategies on the location of the BAO peak. We use the public \texttt{picca}\footnote{https://github.com/igmhub/picca} package for the first two steps and the public \texttt{Vega}\footnote{https://github.com/andreicuceu/vega} package for the third.

\subsection{Continuum Fitting} \label{sec:cont}

We measure the \lya forest flux overdensity field following the exact same procedure as described by \cite{DESI2024.IV.KP6} for the DESI DR1 data and as described by \cite{KP6s6-Cuceu} for the DESI DR1 mocks, with the exception of the BAL masking strategy. We implement differences in the BAL masking strategy through changes to the input quasar catalogs so there are no changes to the codes used for the analysis. The default masking strategy in \texttt{picca} is that all pixels that might be associated with BAL absorption are not included in the analysis, that is they are masked. The quasar catalogs include AI (and BI) values for every BAL, along with the velocity limits for each trough based on the \ion{C}{IV} emission feature. \texttt{picca} reads this information from the catalog and identifies corresponding wavelength range that corresponds to the same velocity offsets associated with emission features that could contaminate the forest region.

We create alternative catalogs for each masking option by setting the AI and velocity ranges for the BAL features to zero for the quasars that we do not want to mask. For example, for our option where we only mask BALs with $AI > 146$, we create a version of the catalog where we set the AI and velocity range values for all BALs with $AI \leq 146$ to be equal to zero. For the case where we only mask BALs with $BI>0$, we set AI and the velocity ranges equal to zero for every BAL that has $BI = 0$. Note the masking for the BALs that remain are based on the velocity range associated with the $AI$ criterion, although these largely correspond to the same pixels as the velocity range for their troughs that meet the BI criterion. In addition, we consider one case where we adjust the redshifts for the BALs as described in Section~\ref{sec:redshifts}. After we adjusted the redshifts, we reran the BAL detection algorithm based on new redshifts. This led to very minor changes in the final catalog of BAL properties, such as an 0.1\% change in the BAL fraction, and consequently very minor changes in which pixels were masked. We refer to this case as ``$AI > 0$ (masked z).'' As a reference, we also perform all of our analysis with no BALs masked (``no masking''), although this is just intended as a point of comparison and not a viable alternative strategy. In addition, DLAs are still masked in all of these options. In total, this corresponds to 19 different options: seven different options with DESI data (including the baseline analysis, no masking, and the updated redshifts) and 12 different options on mocks (six with the true BAL catalog, six with the ``observed'' BAL catalog). 

The \texttt{picca} package uses the catalog information to calculate the \lya forest flux density field in the spectrum of each quasar $q$ as:
\begin{equation}
    \delta_q (\lambda) = \frac{f_q(\lambda)}{\bar{F} (\lambda) C_q(\lambda)} - 1
\end{equation}
where $f_q$ is the measured flux, $\bar{F} (\lambda)$ is the mean transmission of the intergalactic medium, and $C_q$ is the mean quasar continuum. Since the forest is too dense at these redshifts to measure the true continuum directly, the package iteratively fits the quasar continuum and $\bar{F}$ from the data:
\begin{equation}
    \bar{F}(\lambda) C_q(\lambda) = \bar{C} (\lambda_{RF}) \left( a_q + b_q \frac{\Lambda - \Lambda_{min}}{\Lambda_{max} - \Lambda_{min}}\right), 
\end{equation}
where $\bar{C}$ is the mean quasar continuum calculated from the full sample, $\lambda_{RF}$ is the rest-frame wavelength, $\Lambda \equiv \log \lambda$, and $a_q,b_q$ are parameters that are fit separately to each quasar to account for spectral diversity. For more details, see \cite{Ramirez2024}. 

The continua are extremely similar for the different masking options described at the beginning of this section. The ratio of the continuum for each masking option relative to the baseline is only a few tenths of a percent for \lya region A ($1040-1205$\,\AA) and under a percent for \lya region B ($920-1020$\,\AA). We refer to these two regions as Ly$\alpha$(A) and Ly$\alpha$(B). Unsurprisingly, the biggest change is in comparison to the ``no masking'' case. Yet even in this case the dispersion in Ly$\alpha$(B) is slightly less than a percent and about half a percent in Ly$\alpha$(A). These differences are much smaller than those shown in figures 2 -- 4 of \cite{Ennesser2022}, as the figures in that paper just showed the continua for the different subsets of BAL quasars. 

\begin{figure}[htbp]
\centering
\includegraphics[width=1.0\textwidth]{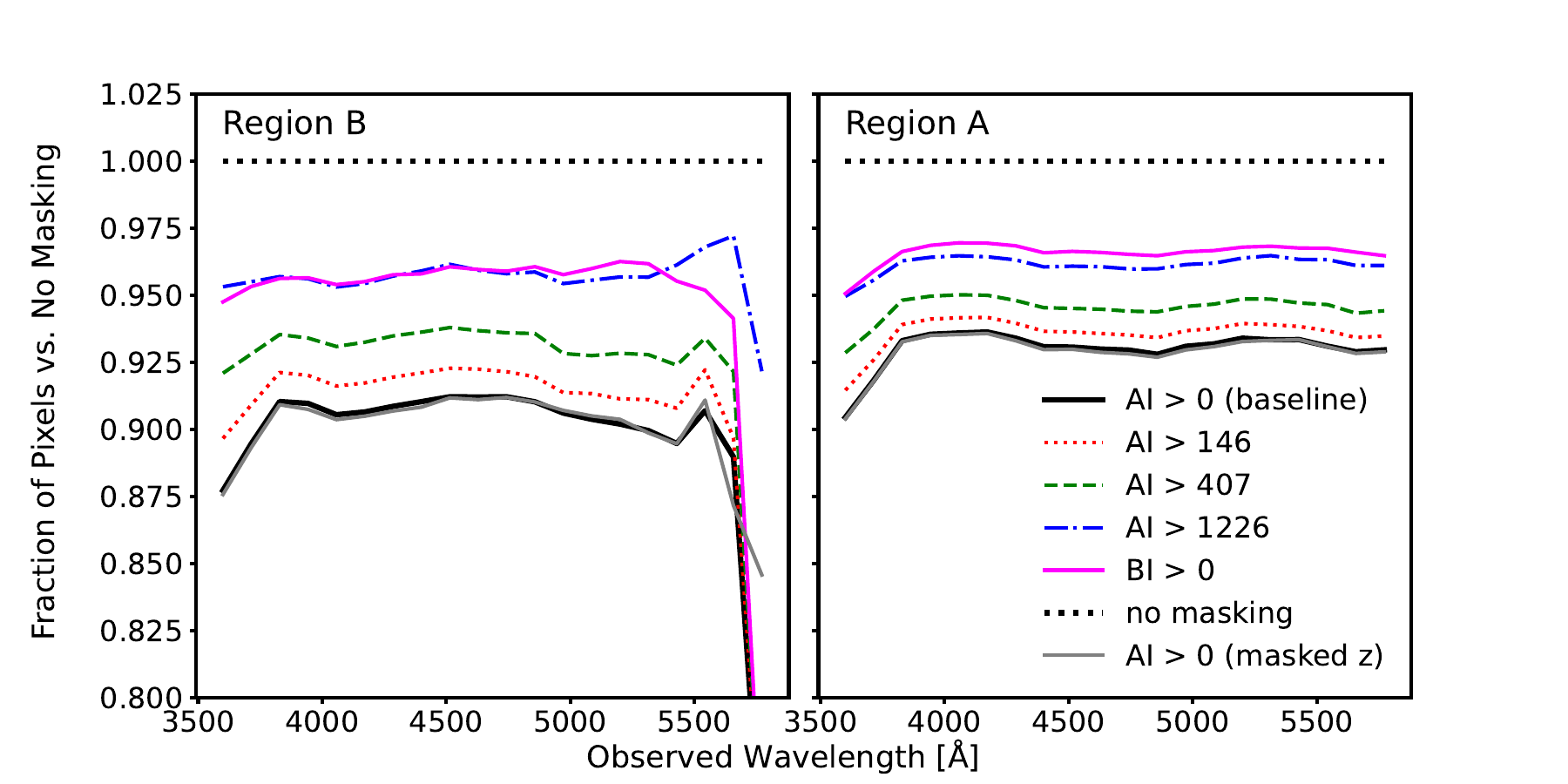}
\caption{Fraction of pixels masked for each option relative to the case with no pixels masked. The typical masked fraction in the baseline analysis is 91\% for Ly$\alpha$(B) ({\it left}, $920-1020$\,\AA) and 93\% for Ly$\alpha$(A) ({\it right}, $1040-1205$\,\AA). There are relatively few analysis pixels at observed wavelengths $> 5500$\,\AA\ in Ly$\alpha$(B) due to the rapid drop in the number of very high redshift quasars. See \S\ref{sec:results} for a description of these masking options. \label{fig:npix}}
\end{figure}

Figure~\ref{fig:npix} shows a plot of the number of pixels that are retained for analysis for each of the masking options relative to the ``no masking'' case. This figure shows that nearly 10\% of pixels in Ly$\alpha$(B) are not included in the baseline analysis, and about 7\% in Ly$\alpha$(A). In contrast, about 5\% are not included when only the most extreme (largest AI or BI-only) cases are masked. The reason the fraction of pixels changes by only a factor of two after eliminating 75\% of the BALs ($AI > 1226$ corresponds to only masking the top quartile) is due to two factors: 1) There is not a one-to-one connection between AI and the number of pixels that are masked, as AI accounts for both the depth and the width of a feature; 2) If enough pixels are masked, then there may be too few pixels remaining in a given forest for it to be retained in the analysis. 


\subsection{Correlation Function and Uncertainties} \label{sec:cf}

We measured all four correlations based on the overdensity fields for each of the nineteen different scenarios described in the previous subsection, that is the seven on data listed in Table~\ref{tab:var} and 12 on mocks (six with the truth catalog, six with the catalog from the detection algorithm). The four correlations are the autocorrelation of Ly$\alpha$(A)$\times$Ly$\alpha$(A), the autocorrelation of Ly$\alpha$(A)$\times$Ly$\alpha$(B), the cross-correlation of Ly$\alpha$(A)$\times$QSO, and the cross correlation of Ly$\alpha$(B)$\times$QSO. Our analysis of the data closely followed the procedure described in detail in \cite{DESI2024.IV.KP6} and our analysis of the mocks closely followed \cite{KP6s6-Cuceu}. 

Very briefly, we use \texttt{picca} to compute these correlations on a spatial grid in comoving separation that extends along $r_{||}$ and across $r_{\perp}$ the line of sight. We convert redshift and angular separations to spatial coordinates with a fiducial cosmology \cite[Planck18, see][]{Planck2018} such that:
\begin{equation}
    r_{||} = [D_c(z_i) - D_c(z_j)] \cos \frac{\Delta \theta}{2},
\end{equation}
\begin{equation}
    r_{\perp} = [D_M(z_i) - D_M(z_j)] \sin \frac{\Delta \theta}{2}. 
\end{equation}
The quantities $z_i$,$z_j$ are the redshifts of the centers of the two bins, $\Delta \theta$ is their separation, $D_c$ is the comoving distance, and $D_M$ is the transverse distance. The bin size for the correlation functions is $4\,h^{-1}$\,Mpc, and we use a $0$ to $200\,h^{-1}$\,Mpc for the auto-correlation, and $-200$ to $200\,h^{-1}$\,Mpc for the cross-correlation. 

The correlation function calculation employs a weighted pair-counting algorithm developed in many previous analyses \cite[e.g.][]{Bautista2017,dMdB2020}. This is:
\begin{equation}
    \xi_M = \frac{\sum_{i,j\in M} w_i w_j \delta_i \delta_j}{\sum_{i,j\in M} w_i w_j}
\end{equation}
for some bin M, where $\delta$ is defined in section~\ref{sec:cont} for the forest. The weights $w_i, w_j$ for the forest account for redshift evolution and pipeline noise, while a separate weight $w_q$ is used for quasars, where this weight includes a model for the evolution of their clustering. 



\begin{table}[htbp]
    \centering
\begin{tabular}{l|c|c|c|c}
    \hline
Masking Option 		 & Ly$\alpha$(A)$\times$Ly$\alpha$(A) & Ly$\alpha$(A)$\times$Ly$\alpha$(B) & Ly$\alpha$(A)$\times$QSO & Ly$\alpha$(B)$\times$QSO \\ 
     \hline
$AI > 0$ (baseline) 		 &  1.000 & 1.000 & 1.000 & 1.000 \\
$AI > 146$ 		 &  0.996 & 0.990 & 0.996 & 0.992 \\
$AI > 407$ 		 &  0.988 & 0.978 & 0.991 & 0.982 \\
$AI > 1226$ 		 &  0.976 & 0.967 & 0.981 & 0.974 \\
$BI > 0$ 		 &  0.976 & 0.973 & 0.982 & 0.977 \\
no masking 		 &  0.988 & 0.997 & 0.972 & 0.983 \\
$AI > 0$ (masked z) 		 &  1.005 & 1.008 & 1.002 & 1.003 \\
     \hline
\end{tabular}
    \caption{Fractional change in the average variance in the four correlations between 80 and 120 $h^{-1}$\,Mpc for the different masking options relative to the baseline. The variance is generally lower when fewer pixels are masked, with the exception that the no masking case has higher variance than some of the options that mask a relatively smaller number of pixels, which may be due to the impact of the BALs on the variance.}
    \label{tab:var}
\end{table}

The key factor in capturing the relative change in the correlation function is how the uncertainties vary for the different masking options in the vicinity of the BAO peak. We calculated the average variance in the range $80\,h^{-1}\,\mathrm{Mpc} < r < 120\,h^{-1}\,\mathrm{Mpc}$ for all four correlation functions and list the fractional change in the variance in table~\ref{tab:var}. This table shows that the variance is smaller for options that mask a smaller portion of BALs, as expected since these options include more of the data, although the change in the variance is only a few percent, which is somewhat smaller than the change in the number of pixels (figure~\ref{fig:npix}). The two main contributors to this trend are likely the relative completeness for different values of AI, in the sense that no longer masking BALs with $0 < AI \leq 146$ are not masking higher SNR pixels, as such weak BALs are only detected in higher SNR spectra \cite{Ennesser2022}. Yet there is a competing, alternative effect that the pixels with BAL absorption will be lower flux and thus lower weight. The ``no masking'' case has lower variance as well, although not as low as some of the least strict masking options, which suggests that unmasked BAL absorption is contributing variance in the ``no masking'' case. Lastly, there is a $<1$\% increase in the variance in the baseline analysis with the new redshifts. This may be because the BAL finder is more successful at finding BALs with the updated redshifts, although the number of BALs is 0.1\% lower (see table~\ref{tab:balfrac}). 





\subsection{Baryon Acoustic Oscillations} \label{sec:bao}

The correlation functions are distorted relative to the true correlation functions by the continuum fitting process described in section~\ref{sec:cf}. This distortion arises because the mean and slope of the continuum is set to zero when we fit for the $a_q$ and $b_q$ parameters, and this removes some large-scale structure information in addition to accounting for the intrinsic diversity of quasars. The analysis for DESI DR1 uses the same approach developed in earlier work \cite{Bautista2017}, which is to build projection matrices to account for this distortion. We use this formalism to forward model the projection matrices for each correlation function into distortion matrices. One change for DESI DR1 is that we generate distortion matrices that are a factor of two higher resolution than the data, namely $2\,h^{-1}\,\mathrm{Mpc}$ rather than $4\,h^{-1}\,\mathrm{Mpc}$. Given the very minor difference in the mean continuum shape and in the total number of pixels, we use the same distortion matrix as the baseline analysis for all of the masking options, rather than compute separate distortion matrices for each option. 

The model fits to the correlation function use the template formalism developed by \cite{Kirkby2013}. This approach splits the isotropic linear power spectrum into a peak and a smooth component, and these components are the templates for the fit. We then add the Kaiser term \cite{Kaiser1987}, models that account for non-linearities, metal absorption, high column density systems, and some other effects and contaminants \cite[see][]{KP6s6-Cuceu}. We do not add BALs to this model, as we mask them before we calculate the correlation function. All of the additional effects and contaminants are added to the smooth and peak components separately and then combined in the fit to the correlation function. The fit varies the coordinates of the BAO feature ($r_{||}, r_{\perp}$) in the template for the peak component with two scale parameters that capture any difference in the values of these coordinates relative to the fiducial model, that is $\alpha_{||} = r_{||}/r_{||,fid}$ and $\alpha_{\perp} = r_{\perp}/r_{\perp,fid}$. We calculate these models and fit them to the data with the \texttt{Vega} package. 

\begin{figure}[htbp]
\centering
\includegraphics[width=1.0\textwidth]{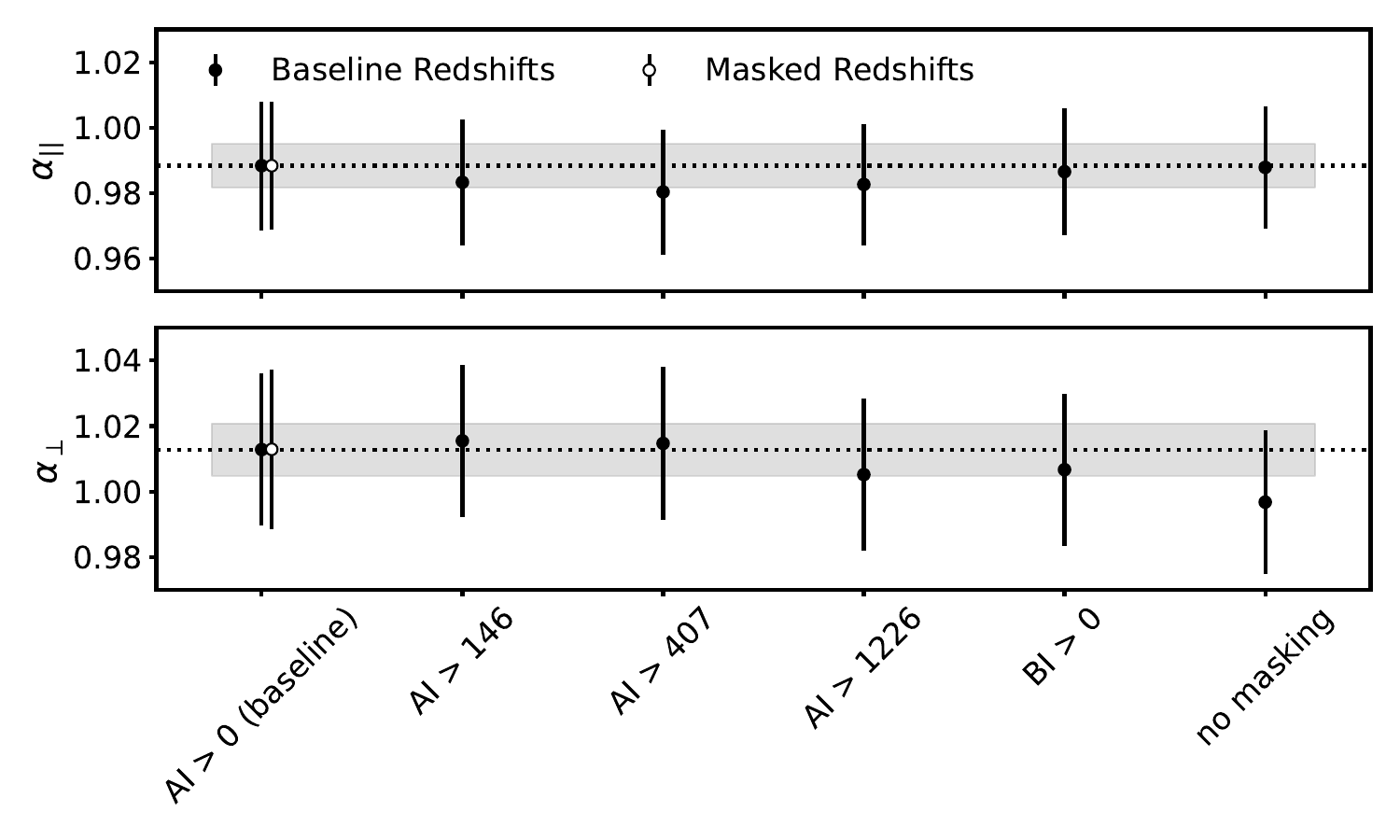}
\caption{BAO parameters $\alpha_{||}$ and $\alpha_{\perp}$ for different masking options on the DESI DR1 data. The baseline redshifts {\it solid circles} are those used for the KP6 analysis \cite{DESI2024.IV.KP6} and the additional masked redshifts points {\it open circles} are the result of rerunning the baseline analysis with the updated redshifts after masking the BAL features. The gray region corresponds to one-third of the size of the uncertainty expected for DESI DR1 analysis. This was used as the threshold for validation prior to unblinding, as described in \cite{DESI2024.IV.KP6}. That paper also notes that these two parameters are correlated with correlation coefficient $\rho = -0.48$. \label{fig:avalsdata}}
\end{figure}

\begin{table}[htbp]
    \centering
\begin{tabular}{l|c|c|c|c}
    \hline
Masking Option & $\alpha_{||}$ & $\sigma_{\alpha_{||}}$ & $\alpha_{\perp}$ & $\sigma_{\alpha_{\perp}}$ \\
    \hline
$AI > 0$ (baseline)      & 0.988 & 0.020 & 1.013 & 0.023 \\
$AI > 146$       & 0.983 & 0.019 & 1.015 & 0.023 \\
$AI > 407$       & 0.980 & 0.019 & 1.015 & 0.023 \\
$AI > 1226$      & 0.983 & 0.019 & 1.005 & 0.023 \\
$BI > 0$         & 0.987 & 0.019 & 1.007 & 0.023 \\
no masking     & 0.988 & 0.019 & 0.997 & 0.022 \\
$AI > 0$ (masked z)      & 0.988 & 0.020 & 1.013 & 0.024 \\
     \hline
\end{tabular}
    \caption{Best fit BAL parameters and uncertainties for the masking options applied to data, as well as the extreme case with no BAL masking, and the results with updated redshifts for the BALs after masking their troughs. }
    \label{tab:avalsdata}
\end{table}

The model fits to data and the mocks include many parameters in addition to the $\alpha_{||}$ and $\alpha_{\perp}$ values that provide the BAO peak location relative to the model. These parameters include separate bias parameters for the \lya forest, high column density (HCD) systems, quasars, and several metal lines, redshift space distortion parameters for the \lya forest and HCDs, and a model for the column density distribution of the HCDs. HCDs represent absorption systems with $\log N_{\ion{H}{I}} < 20.3$ that are below the threshold for damped \lya systems, which are masked in the analysis, and yet higher column density than the forest. There are also model parameters for statistical quasar redshift errors, quasar non-linear velocities, and smoothing terms to the models for mocks that take into account the simulation grid cell size. All of these analysis steps are the same as those applied to the DESI DR1 data analysis \cite{DESI2024.IV.KP6} and mocks \cite{KP6s6-Cuceu} and we refer to those papers for more details. 

We modeled all four correlations for each of the BAL masking options with \texttt{Vega} and calculated $\alpha_{||}$ and $\alpha_{\perp}$, in addition to the other model parameters. Figure~\ref{fig:avalsdata} shows the variation in the values for the different masking options and the numerical values are listed in table~\ref{tab:avalsdata}. The horizontal, dotted line in each panel indicates the values of the two parameters in the $AI > 0$ baseline masking option, and is also identical to the value presented in the DESI DR1 Key Paper on the \lya forest \cite{DESI2024.IV.KP6}. Figure~\ref{fig:avalsdata} shows the main result of our study, namely that the measurements of the BAO parameters are quite insensitive to a broad range of reasonable masking decisions, even the most extreme case of only masking BALs based on the $BI > 0$ criterion. The variation between the measured values of $\alpha$ for these options are more than a factor of five smaller than the uncertainties in the measurements. Redshift errors have a similarly minimal impact. The model fit based on the updated redshifts after masking the BAL features is nearly identical to the baseline model, and we therefore did not explore other masking options with these new redshifts. We also note that some variation is expected with larger changes in the number of (un)masked pixels because of the change in the data (e.g. see figure~\ref{fig:npix}). The most significant change is the extreme and unreasonable case where no BALs are masked. 

\begin{figure}[htbp]
\centering
\includegraphics[width=1.0\textwidth]{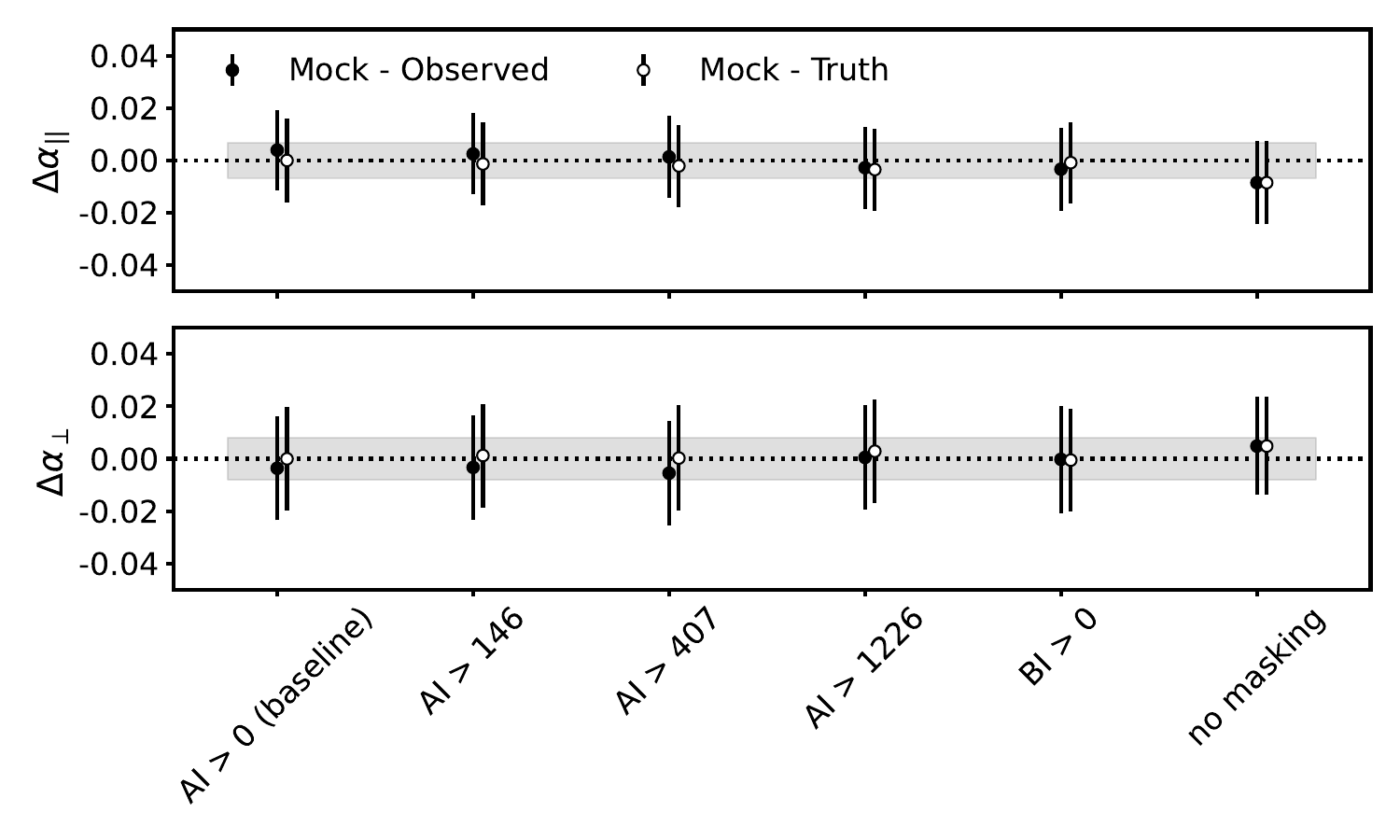}
\caption{Same as figure~\ref{fig:avalsdata} for the BAO parameters $\alpha_{||}$ and $\alpha_{\perp}$ for different masking options on the DESI DR1 mocks. \label{fig:avalsmocks}}
\end{figure}

\begin{table}[htbp]
    \centering
\begin{tabular}{l|c|c|c|c|c|c|c|c}
    \hline
Masking Option & $\Delta \alpha_{||}$ & $\sigma_{\alpha_{||}}$ & $\Delta \alpha_{t,||}$ & $\sigma_{\alpha_{t,||}}$ & $\Delta \alpha_{\perp}$ & $\sigma_{\alpha_{\perp}}$ & $\Delta \alpha_{t,\perp}$ & $\sigma_{t,\alpha_{\perp}}$ \\
\hline
$AI > 0$ (baseline) 	 & 0.004 & 0.015 & 0.000 & 0.016 & -0.004 & 0.020 & 0.000 & 0.020 \\
$AI > 146$ 	 & 0.002 & 0.015 & -0.001 & 0.016 & -0.003 & 0.020 & 0.001 & 0.020 \\
$AI > 407$ 	 & 0.001 & 0.016 & -0.002 & 0.016 & -0.005 & 0.020 & 0.000 & 0.020 \\
$AI > 1226$ 	 & -0.003 & 0.016 & -0.003 & 0.016 & 0.001 & 0.020 & 0.003 & 0.020 \\
$BI > 0$ 	 & -0.003 & 0.016 & -0.001 & 0.015 & -0.000 & 0.020 & -0.001 & 0.019 \\
no masking 	 & -0.008 & 0.016 & -0.008 & 0.016 & 0.005 & 0.019 & 0.005 & 0.019 \\
     \hline
\end{tabular}
    \caption{Best fit BAL parameters and uncertainties for the masking options applied to mocks, as well as the extreme case with no BAL masking. The values of $\Delta \alpha_{||}$ and $\Delta \alpha_{\perp}$ are the difference between a given masking option and the baseline result with the BAL truth catalog. The columns labeled $\Delta \alpha_{t,||}$ and $\Delta \alpha_{t,\perp}$ are calculated with the BAL truth catalog and the other columns are from catalogs based on the BAL identification algorithm. }
    \label{tab:avalsmocks}
\end{table}
We also modeled all four correlations with the mock data to study differences due to completeness, and therefore we fit models for each masking option based on both the truth catalog of BALs and the BAL catalog recovered with the identification algorithm. The results for each masking option are shown in figure~\ref{fig:avalsmocks}, where there is a separate point for the observed mock and the true mock catalog for each masking option. The variations between the fits to the observed and truth mock data are $\sim 0.003$ for both parameters, or about seven times smaller than the uncertainties in the DESI DR1 data, which demonstrates that incompleteness in the BAL identification does not impact the BAO fits. In addition, the variation for different masking options for these mock datasets is similar in magnitude to the variations we observed with the observational data shown in figure~\ref{fig:avalsdata}. We therefore conclude that both incompleteness in the BAL identification algorithm and modest variations in the BAL masking strategy do not impact the BAO parameters at a level that would impact the DESI DR1 results. 

\section{Summary} \label{sec:summary}

BAL quasars introduce systematic errors into the use of the \lya forest for cosmology studies because they add absorption in the forest region that is unrelated to the IGM and they add measurement errors to the quasars that trace large scale structure. The baseline strategy adopted for DESI DR1 analysis is to use the velocity range of BAL troughs with $AI > 0$ in the region around \ion{C}{IV} to identify pixels in the \lya forest that may be affected by BAL absorption from other emission features. This strategy is motivated by the work of \cite{Ennesser2022} with eBOSS data, and first applied to DESI with the EDR study by \cite{Filbert2023}. A careful study of mock data by \cite{Garcia2023} showed that BALs could introduce redshift errors, and showed that masking the BAL features reduced the redshift errors of the BAL quasars to be comparable to quasars that do not show BAL features. Subsequent work by \cite{Filbert2023} applied this strategy to DESI EDR data and then work by \cite{Brodzeller2023} and \cite{KP6s4-Bault} confirmed the decrease in systematic and random redshift errors with cross-correlation studies. 

We have studied a range of alternative masking strategies with the DESI DR1 dataset to quantify the impact of changes in the masking strategy to the BAO measurements. These masking strategies explore not masking progressively stronger BAL features, as parameterized by the AI parameter. Specifically, we performed a complete end-to-end analysis with the lowest AI quartile unmasked, the lowest half of the AI distribution unmasked, all but the largest quartile unmasked, and only masking quasars that met the BI criterion. In all cases the BAO parameters change by less than a percent, and we therefore do not expect they will impact \lya BAO measurements even with the final DESI dataset. The change is also negligible when we recompute the redshifts for the BALs after masking their absorption and then performing the complete analysis. There is a minor change in the parameters when we do not mask the BAL features at all, which is not a reasonable alternative to the baseline analysis, yet this case is representative of the largest potential impact of the BALs on this measurement. The variations in the two $\alpha$ values for different masking options on the observational data are shown in figure~\ref{fig:avalsdata}.

The BAL identification algorithm does not recover all BALs in the data, and particularly suffers from incompleteness in low SNR data. This was first pointed out in the analysis of BAL masking by \cite{Ennesser2022}, who showed that eliminating BALs from the analysis preferentially eliminated the highest SNR spectra, and then studied with DESI EDR data by \cite{Filbert2023}. In this work we have measured the completeness of the BAL identification algorithm as a function of SNR and AI value with a study of ten of the DESI DR1 mock datasets produced by \cite{KP6s6-Cuceu}. The cumulative completeness is 60\% for all BALs, although higher for BALs with larger AI values (see figure~\ref{fig:cp} and table~\ref{tab:cp}). We expect it is somewhat lower for the data, as the mocks have somewhat higher SNR (see section~\ref{sec:comp}) 

The ultimate applicability of our mock results to data depends on the fidelity of the BALs in the mocks. We measured the distribution of several BAL parameters and found that the BALs in the mocks have somewhat fewer BALs based on the AI criterion, are a good match in number with respect to the BI criterion, and agree well with respect to the velocity ranges of the BAL troughs, with the exception of BALs on the blue wing of the \ion{C}{IV} emission feature. These differences between the mocks and the DESI data may be because the BAL templates used for the mocks were tuned to match the properties of BALs in SDSS. One clear area for improvement of the BAL templates is to use the DESI data to construct new templates that are a better match to DESI. It may also be worthwhile to explore options to improve the performance of the identification algorithm in the immediate vicinity of the \ion{C}{IV} emission feature, although this may not be readily tractable because this line often appears asymmetric even in quasars that do not show BAL features. 

While there are some limitations in the fidelity of the mocks with respect to BALs in DESI observations, we find negligible differences in the BAO fits between the truth catalog and the catalog from the BAL identification algorithm for all of the masking options (see figure~\ref{fig:avalsmocks}). This indicates the completeness of the catalogs has a correspondingly negligible impact on the BAO results. Furthermore, the differences between the two mock catalogs for each option approximately span the range of variation between the BAL templates and the DESI data, and therefore there is no indication that the fidelity of the mocks impacts the BAO results. 

The main motivation for the future development of the BAL templates will likely be other cosmological analysis with DESI data. The so-called ``full shape'' analysis developed by \cite{Cuceu2021} and then applied to eBOSS mocks and data \cite{Cuceu2023a,Cuceu2023b} has promise to improve the cosmological parameter estimation from DESI \lya forest data by up to a factor of two relative to the BAO peak alone. This analysis is called full shape because it uses information in the correlation function across most spatial scales, and not just those in the vicinity of the BAO peak. Yet this use of more scales makes it more important to model and mitigate sources of systematic errors such as BALs. BALs may also impact the measurement of the one-dimensional power spectrum ($P_{1D}$) of the \lya forest. While BALs have been masked in the EDR analysis by \cite{Karacayli2024} with the optimal quadratic estimator, this is less straightforward with the Fourier transform approach \cite{Ravoux2023}, and the full impact of BALs on the future DESI DR1 $P_{1D}$ \lya results is an area of ongoing study. 




\appendix
\section{Completeness and Purity Data}
\label{sec:cpdata}

This appendix contains the tabulated data discussed in section~\ref{sec:comp}. Table~\ref{tab:cp} lists the differential and cumulative completeness and purity as a function of SNR based on ten mock datasets. 

\begin{table}[htbp]
    \centering
\begin{tabular}{r|c|c|c|c|c|c}
    \hline
SNR      & Data Fraction & Mock Fraction  & Complete & Purity & Complete & Purity \\
         & (Differential)  & (Differential)   & (Differential) & (Differential) & (Cumulative) & (Cumulative) \\ 
     \hline
     \hline
SNR 	 & Data 	 & Mock 	 & Completeness & Purity & Completeness & Purity \\
 0.5 	 & 0.306 	 & 0.210 	 & 0.000 	 & 1.000 	 & 0.000 	 & 1.000  \\
 1.5 	 & 0.258 	 & 0.293 	 & 0.569 	 & 0.900 	 & 0.332 	 & 0.942  \\
 2.5 	 & 0.140 	 & 0.160 	 & 0.790 	 & 0.923 	 & 0.443 	 & 0.937  \\
 3.5 	 & 0.084 	 & 0.096 	 & 0.868 	 & 0.908 	 & 0.497 	 & 0.934  \\
 4.5 	 & 0.056 	 & 0.063 	 & 0.907 	 & 0.892 	 & 0.528 	 & 0.930  \\
 5.5 	 & 0.039 	 & 0.044 	 & 0.932 	 & 0.864 	 & 0.549 	 & 0.927  \\
 6.5 	 & 0.028 	 & 0.031 	 & 0.945 	 & 0.841 	 & 0.562 	 & 0.924  \\
 7.5 	 & 0.021 	 & 0.023 	 & 0.962 	 & 0.809 	 & 0.573 	 & 0.921  \\
 8.5 	 & 0.015 	 & 0.018 	 & 0.971 	 & 0.791 	 & 0.580 	 & 0.919  \\
 9.5 	 & 0.012 	 & 0.014 	 & 0.973 	 & 0.757 	 & 0.586 	 & 0.916  \\
 10.5 	 & 0.009 	 & 0.010 	 & 0.969 	 & 0.739 	 & 0.590 	 & 0.915  \\
 11.5 	 & 0.007 	 & 0.008 	 & 0.983 	 & 0.718 	 & 0.593 	 & 0.913  \\
 12.5 	 & 0.006 	 & 0.007 	 & 0.978 	 & 0.689 	 & 0.596 	 & 0.911  \\
 13.5 	 & 0.005 	 & 0.005 	 & 0.978 	 & 0.647 	 & 0.598 	 & 0.910  \\
 14.5 	 & 0.004 	 & 0.004 	 & 0.982 	 & 0.661 	 & 0.599 	 & 0.909  \\
 15.5 	 & 0.003 	 & 0.004 	 & 0.982 	 & 0.594 	 & 0.601 	 & 0.908  \\
 16.5 	 & 0.002 	 & 0.003 	 & 0.989 	 & 0.572 	 & 0.602 	 & 0.907  \\
 17.5 	 & 0.002 	 & 0.002 	 & 0.987 	 & 0.584 	 & 0.603 	 & 0.906  \\
 18.5 	 & 0.002 	 & 0.002 	 & 0.988 	 & 0.525 	 & 0.604 	 & 0.905  \\
 19.5 	 & 0.001 	 & 0.002 	 & 0.988 	 & 0.520 	 & 0.604 	 & 0.905  \\
 \hline
\end{tabular}
    \caption{Completeness and Purity of the BAL identification algorithm as a function of SNR in the \ion{C}{IV} region. The SNR values in the first column represent the centers of each bin. The second and third columns shows the differential fraction of the data and mock quasar catalogs that have SNR less than or equal to the SNR bin. The differential completeness and purity are listed in the fourth and fifth columns, and the cumulative completeness and purity are in the last two columns. The average completeness and purity are 60\% and 91\%, respectively, based on the average of ten mock datasets. The dispersion between mocks is illustrated with a different line for each mock in Figure~\ref{fig:cp}.}
    \label{tab:cp}
\end{table}

\section{Data Availability}
The data used in this work will be made public as part of DESI Data Release 1 (details at \url{https://data.desi.lbl.gov/doc/releases/}). The data points corresponding to the figures
are available at \url{https://zenodo.org/records/11194879}. 

\section{Author Affiliations}
\label{sec:affiliations}

\noindent \hangindent=.5cm $^{1}${Department of Astronomy, The Ohio State University, 4055 McPherson Laboratory, 140 W 18th Avenue, Columbus, OH 43210, USA}

\noindent \hangindent=.5cm $^{2}${Center for Cosmology and AstroParticle Physics, The Ohio State University, 191 West Woodruff Avenue, Columbus, OH 43210, USA}

\noindent \hangindent=.5cm $^{3}${Department of Physics, The Ohio State University, 191 West Woodruff Avenue, Columbus, OH 43210, USA}

\noindent \hangindent=.5cm $^{4}${Department of Physics and Astronomy, The University of Utah, 115 South 1400 East, Salt Lake City, UT 84112, USA}

\noindent \hangindent=.5cm $^{5}${Lawrence Berkeley National Laboratory, 1 Cyclotron Road, Berkeley, CA 94720, USA}

\noindent \hangindent=.5cm $^{6}${Physics Dept., Boston University, 590 Commonwealth Avenue, Boston, MA 02215, USA}

\noindent \hangindent=.5cm $^{7}${Department of Physics \& Astronomy, University College London, Gower Street, London, WC1E 6BT, UK}

\noindent \hangindent=.5cm $^{8}${Instituto de F\'{\i}sica, Universidad Nacional Aut\'{o}noma de M\'{e}xico,  Cd. de M\'{e}xico  C.P. 04510,  M\'{e}xico}

\noindent \hangindent=.5cm $^{9}${NSF NOIRLab, 950 N. Cherry Ave., Tucson, AZ 85719, USA}

\noindent \hangindent=.5cm $^{10}${Departamento de F\'isica, Universidad de los Andes, Cra. 1 No. 18A-10, Edificio Ip, CP 111711, Bogot\'a, Colombia}

\noindent \hangindent=.5cm $^{11}${Observatorio Astron\'omico, Universidad de los Andes, Cra. 1 No. 18A-10, Edificio H, CP 111711 Bogot\'a, Colombia}

\noindent \hangindent=.5cm $^{12}${Institut d'Estudis Espacials de Catalunya (IEEC), 08034 Barcelona, Spain}

\noindent \hangindent=.5cm $^{13}${Institute of Cosmology and Gravitation, University of Portsmouth, Dennis Sciama Building, Portsmouth, PO1 3FX, UK}

\noindent \hangindent=.5cm $^{14}${Institute of Space Sciences, ICE-CSIC, Campus UAB, Carrer de Can Magrans s/n, 08913 Bellaterra, Barcelona, Spain}

\noindent \hangindent=.5cm $^{15}${Departamento de F\'{i}sica, Universidad de Guanajuato - DCI, C.P. 37150, Leon, Guanajuato, M\'{e}xico}

\noindent \hangindent=.5cm $^{16}${The Ohio State University, Columbus, 43210 OH, USA}

\noindent \hangindent=.5cm $^{17}${Sorbonne Universit\'{e}, CNRS/IN2P3, Laboratoire de Physique Nucl\'{e}aire et de Hautes Energies (LPNHE), FR-75005 Paris, France}

\noindent \hangindent=.5cm $^{18}${Departament de F\'{i}sica, Serra H\'{u}nter, Universitat Aut\`{o}noma de Barcelona, 08193 Bellaterra (Barcelona), Spain}

\noindent \hangindent=.5cm $^{19}${Institut de F\'{i}sica d’Altes Energies (IFAE), The Barcelona Institute of Science and Technology, Campus UAB, 08193 Bellaterra Barcelona, Spain}

\noindent \hangindent=.5cm $^{20}${Instituci\'{o} Catalana de Recerca i Estudis Avan\c{c}ats, Passeig de Llu\'{\i}s Companys, 23, 08010 Barcelona, Spain}

\noindent \hangindent=.5cm $^{21}${Department of Astronomy, Tsinghua University, 30 Shuangqing Road, Haidian District, Beijing, China, 100190}

\noindent \hangindent=.5cm $^{22}${Department of Physics and Astronomy, Siena College, 515 Loudon Road, Loudonville, NY 12211, USA}

\noindent \hangindent=.5cm $^{23}${Instituto Avanzado de Cosmolog\'{\i}a A.~C., San Marcos 11 - Atenas 202. Magdalena Contreras, 10720. Ciudad de M\'{e}xico, M\'{e}xico}

\noindent \hangindent=.5cm $^{24}${IRFU, CEA, Universit\'{e} Paris-Saclay, F-91191 Gif-sur-Yvette, France}

\noindent \hangindent=.5cm $^{25}${Department of Physics and Astronomy, University of Waterloo, 200 University Ave W, Waterloo, ON N2L 3G1, Canada}

\noindent \hangindent=.5cm $^{26}${Perimeter Institute for Theoretical Physics, 31 Caroline St. North, Waterloo, ON N2L 2Y5, Canada}

\noindent \hangindent=.5cm $^{27}${Waterloo Centre for Astrophysics, University of Waterloo, 200 University Ave W, Waterloo, ON N2L 3G1, Canada}

\noindent \hangindent=.5cm $^{28}${Departament de F\'isica, EEBE, Universitat Polit\`ecnica de Catalunya, c/Eduard Maristany 10, 08930 Barcelona, Spain}

\noindent \hangindent=.5cm $^{29}${Space Sciences Laboratory, University of California, Berkeley, 7 Gauss Way, Berkeley, CA  94720, USA}

\noindent \hangindent=.5cm $^{30}${University of California, Berkeley, 110 Sproul Hall \#5800 Berkeley, CA 94720, USA}

\noindent \hangindent=.5cm $^{31}${Instituto de Astrof\'{i}sica de Andaluc\'{i}a (CSIC), Glorieta de la Astronom\'{i}a, s/n, E-18008 Granada, Spain}

\noindent \hangindent=.5cm $^{32}${Aix Marseille Univ, CNRS/IN2P3, CPPM, Marseille, France}

\noindent \hangindent=.5cm $^{33}${Universit\'{e} Clermont-Auvergne, CNRS, LPCA, 63000 Clermont-Ferrand, France}

\noindent \hangindent=.5cm $^{34}${Department of Physics, Kansas State University, 116 Cardwell Hall, Manhattan, KS 66506, USA}

\noindent \hangindent=.5cm $^{35}${Department of Physics and Astronomy, Sejong University, Seoul, 143-747, Korea}

\noindent \hangindent=.5cm $^{36}${CIEMAT, Avenida Complutense 40, E-28040 Madrid, Spain}

\noindent \hangindent=.5cm $^{37}${Department of Physics, University of Michigan, Ann Arbor, MI 48109, USA}

\noindent \hangindent=.5cm $^{38}${University of Michigan, Ann Arbor, MI 48109, USA}

\noindent \hangindent=.5cm $^{39}${Department of Physics \& Astronomy, Ohio University, Athens, OH 45701, USA}

\noindent \hangindent=.5cm $^{40}${Excellence Cluster ORIGINS, Boltzmannstrasse 2, D-85748 Garching, Germany}

\noindent \hangindent=.5cm $^{41}${University Observatory, Faculty of Physics, Ludwig-Maximilians-Universit\"{a}t, Scheinerstr. 1, 81677 M\"{u}nchen, Germany}

\noindent \hangindent=.5cm $^{42}${National Astronomical Observatories, Chinese Academy of Sciences, A20 Datun Rd., Chaoyang District, Beijing, 100012, P.R. China}




\acknowledgments

PM appreciates culinary motivation from XM to complete this work. PM and LE acknowledge support from the United States Department of Energy, Office of High Energy Physics under Award Number DE-SC0011726. AC acknowledges support provided by NASA through the NASA Hubble Fellowship grant HST-HF2-51526.001-A awarded by the Space Telescope Science Institute, which is operated by the Association of Universities for Research in Astronomy, Incorporated, under NASA contract NAS5-26555. 

This material is based upon work supported by the U.S. Department of Energy (DOE), Office of Science, Office of High-Energy Physics, under Contract No. DE–AC02–05CH11231, and by the National Energy Research Scientific Computing Center, a DOE Office of Science User Facility under the same contract. Additional support for DESI was provided by the U.S. National Science Foundation (NSF), Division of Astronomical Sciences under Contract No. AST-0950945 to the NSF’s National Optical-Infrared Astronomy Research Laboratory; the Science and Technology Facilities Council of the United Kingdom; the Gordon and Betty Moore Foundation; the Heising-Simons Foundation; the French Alternative Energies and Atomic Energy Commission (CEA); the National Council of Humanities, Science and Technology of Mexico (CONAHCYT); the Ministry of Science and Innovation of Spain (MICINN), and by the DESI Member Institutions: \url{https://www.desi.lbl.gov/collaborating-institutions}. Any opinions, findings, and conclusions or recommendations expressed in this material are those of the author(s) and do not necessarily reflect the views of the U. S. National Science Foundation, the U. S. Department of Energy, or any of the listed funding agencies.

The authors are honored to be permitted to conduct scientific research on Iolkam Du’ag (Kitt Peak), a mountain with particular significance to the Tohono O’odham Nation.

\bibliographystyle{JHEP.bst}
\bibliography{main-rev}

\providecommand{\href}[2]{#2}\begingroup\raggedright\begin{thebibliography}{10}

\bibitem{Weinberg2013}
D.H.~{Weinberg}, M.J.~{Mortonson}, D.J.~{Eisenstein}, C.~{Hirata}, A.G.~{Riess}
  and E.~{Rozo}, \emph{{Observational probes of cosmic acceleration}},
  \href{https://doi.org/10.1016/j.physrep.2013.05.001}{\emph{\physrep}
  {\bfseries 530} (2013) 87} [\href{https://arxiv.org/abs/1201.2434}{{\ttfamily
  1201.2434}}].

\bibitem{Planck2018}
{Planck Collaboration}, N.~{Aghanim}, Y.~{Akrami}, M.~{Ashdown}, J.~{Aumont},
  C.~{Baccigalupi} et~al., \emph{{Planck 2018 results. VI. Cosmological
  parameters}}, \href{https://doi.org/10.1051/0004-6361/201833910}{\emph{\aap}
  {\bfseries 641} (2020) A6}
  [\href{https://arxiv.org/abs/1807.06209}{{\ttfamily 1807.06209}}].

\bibitem{Dawson2013}
K.S.~{Dawson}, D.J.~{Schlegel}, C.P.~{Ahn}, S.F.~{Anderson}, {\'E}.~{Aubourg},
  S.~{Bailey} et~al., \emph{{The Baryon Oscillation Spectroscopic Survey of
  SDSS-III}}, \href{https://doi.org/10.1088/0004-6256/145/1/10}{\emph{\aj}
  {\bfseries 145} (2013) 10} [\href{https://arxiv.org/abs/1208.0022}{{\ttfamily
  1208.0022}}].

\bibitem{Abbott2018}
T.M.C.~{Abbott}, F.B.~{Abdalla}, A.~{Alarcon}, J.~{Aleksi{\'c}}, S.~{Allam},
  S.~{Allen} et~al., \emph{{Dark Energy Survey year 1 results: Cosmological
  constraints from galaxy clustering and weak lensing}},
  \href{https://doi.org/10.1103/PhysRevD.98.043526}{\emph{\prd} {\bfseries 98}
  (2018) 043526} [\href{https://arxiv.org/abs/1708.01530}{{\ttfamily
  1708.01530}}].

\bibitem{Alam2021}
S.~{Alam}, M.~{Aubert}, S.~{Avila}, C.~{Balland}, J.E.~{Bautista},
  M.A.~{Bershady} et~al., \emph{{Completed SDSS-IV extended Baryon Oscillation
  Spectroscopic Survey: Cosmological implications from two decades of
  spectroscopic surveys at the Apache Point Observatory}},
  \href{https://doi.org/10.1103/PhysRevD.103.083533}{\emph{\prd} {\bfseries
  103} (2021) 083533} [\href{https://arxiv.org/abs/2007.08991}{{\ttfamily
  2007.08991}}].

\bibitem{Levi2013}
M.~{Levi}, C.~{Bebek}, T.~{Beers}, R.~{Blum}, R.~{Cahn}, D.~{Eisenstein}
  et~al., \emph{{The DESI Experiment, a whitepaper for Snowmass 2013}},
  \href{https://doi.org/10.48550/arXiv.1308.0847}{\emph{arXiv e-prints} (2013)
  arXiv:1308.0847} [\href{https://arxiv.org/abs/1308.0847}{{\ttfamily
  1308.0847}}].

\bibitem{DESI2016a.Science}
{DESI Collaboration}, A.~{Aghamousa}, J.~{Aguilar}, S.~{Ahlen}, S.~{Alam},
  L.E.~{Allen} et~al., \emph{{The DESI Experiment Part I: Science,Targeting,
  and Survey Design}}, {\emph{arXiv e-prints} (2016) arXiv:1611.00036}
  [\href{https://arxiv.org/abs/1611.00036}{{\ttfamily 1611.00036}}].

\bibitem{DESI2016b.Instr}
{DESI Collaboration}, A.~{Aghamousa}, J.~{Aguilar}, S.~{Ahlen}, S.~{Alam},
  L.E.~{Allen} et~al., \emph{{The DESI Experiment Part II: Instrument Design}},
  {\emph{arXiv e-prints} (2016) arXiv:1611.00037}
  [\href{https://arxiv.org/abs/1611.00037}{{\ttfamily 1611.00037}}].

\bibitem{Busca2013}
N.G.~{Busca}, T.~{Delubac}, J.~{Rich}, S.~{Bailey}, A.~{Font-Ribera},
  D.~{Kirkby} et~al., \emph{{Baryon acoustic oscillations in the Ly{$\alpha$}
  forest of BOSS quasars}},
  \href{https://doi.org/10.1051/0004-6361/201220724}{\emph{\aap} {\bfseries
  552} (2013) A96} [\href{https://arxiv.org/abs/1211.2616}{{\ttfamily
  1211.2616}}].

\bibitem{Slosar2013}
A.~{Slosar}, V.~{Ir{\v{s}}i{\v{c}}}, D.~{Kirkby}, S.~{Bailey}, N.G.~{Busca},
  T.~{Delubac} et~al., \emph{{Measurement of baryon acoustic oscillations in
  the Lyman-{\ensuremath{\alpha}} forest fluctuations in BOSS data release 9}},
  \href{https://doi.org/10.1088/1475-7516/2013/04/026}{\emph{\jcap} {\bfseries
  2013} (2013) 026} [\href{https://arxiv.org/abs/1301.3459}{{\ttfamily
  1301.3459}}].

\bibitem{Kirkby2013}
D.~{Kirkby}, D.~{Margala}, A.~{Slosar}, S.~{Bailey}, N.G.~{Busca}, T.~{Delubac}
  et~al., \emph{{Fitting methods for baryon acoustic oscillations in the
  Lyman-{$\alpha$} forest fluctuations in BOSS data release 9}},
  \href{https://doi.org/10.1088/1475-7516/2013/03/024}{\emph{\jcap} {\bfseries
  3} (2013) 024} [\href{https://arxiv.org/abs/1301.3456}{{\ttfamily
  1301.3456}}].

\bibitem{Eisenstein2011}
D.J.~{Eisenstein}, D.H.~{Weinberg}, E.~{Agol}, H.~{Aihara}, C.~{Allende
  Prieto}, S.F.~{Anderson} et~al., \emph{{SDSS-III: Massive Spectroscopic
  Surveys of the Distant Universe, the Milky Way, and Extra-Solar Planetary
  Systems}}, \href{https://doi.org/10.1088/0004-6256/142/3/72}{\emph{\aj}
  {\bfseries 142} (2011) 72} [\href{https://arxiv.org/abs/1101.1529}{{\ttfamily
  1101.1529}}].

\bibitem{Ahn2012}
C.P.~{Ahn}, R.~{Alexandroff}, C.~{Allende Prieto}, S.F.~{Anderson},
  T.~{Anderton}, B.H.~{Andrews} et~al., \emph{{The Ninth Data Release of the
  Sloan Digital Sky Survey: First Spectroscopic Data from the SDSS-III Baryon
  Oscillation Spectroscopic Survey}},
  \href{https://doi.org/10.1088/0067-0049/203/2/21}{\emph{\apjs} {\bfseries
  203} (2012) 21} [\href{https://arxiv.org/abs/1207.7137}{{\ttfamily
  1207.7137}}].

\bibitem{FontRibera2013}
A.~{Font-Ribera}, E.~{Arnau}, J.~{Miralda-Escud{\'e}}, E.~{Rollinde},
  J.~{Brinkmann}, J.R.~{Brownstein} et~al., \emph{{The large-scale quasar-Lyman
  {$\alpha$} forest cross-correlation from BOSS}},
  \href{https://doi.org/10.1088/1475-7516/2013/05/018}{\emph{\jcap} {\bfseries
  5} (2013) 018} [\href{https://arxiv.org/abs/1303.1937}{{\ttfamily
  1303.1937}}].

\bibitem{Alam2015}
S.~{Alam}, F.D.~{Albareti}, C.~{Allende Prieto}, F.~{Anders}, S.F.~{Anderson},
  T.~{Anderton} et~al., \emph{{The Eleventh and Twelfth Data Releases of the
  Sloan Digital Sky Survey: Final Data from SDSS-III}},
  \href{https://doi.org/10.1088/0067-0049/219/1/12}{\emph{\apjs} {\bfseries
  219} (2015) 12} [\href{https://arxiv.org/abs/1501.00963}{{\ttfamily
  1501.00963}}].

\bibitem{Ahumada2020}
R.~{Ahumada}, C.~{Allende Prieto}, A.~{Almeida}, F.~{Anders}, S.F.~{Anderson},
  B.H.~{Andrews} et~al., \emph{{The 16th Data Release of the Sloan Digital Sky
  Surveys: First Release from the APOGEE-2 Southern Survey and Full Release of
  eBOSS Spectra}}, \href{https://doi.org/10.3847/1538-4365/ab929e}{\emph{\apjs}
  {\bfseries 249} (2020) 3} [\href{https://arxiv.org/abs/1912.02905}{{\ttfamily
  1912.02905}}].

\bibitem{Delubac2015}
T.~{Delubac}, J.E.~{Bautista}, N.G.~{Busca}, J.~{Rich}, D.~{Kirkby},
  S.~{Bailey} et~al., \emph{{Baryon acoustic oscillations in the
  Ly{\ensuremath{\alpha}} forest of BOSS DR11 quasars}},
  \href{https://doi.org/10.1051/0004-6361/201423969}{\emph{\aap} {\bfseries
  574} (2015) A59} [\href{https://arxiv.org/abs/1404.1801}{{\ttfamily
  1404.1801}}].

\bibitem{Bautista2017}
J.E.~{Bautista}, N.G.~{Busca}, J.~{Guy}, J.~{Rich}, M.~{Blomqvist}, H.~{du Mas
  des Bourboux} et~al., \emph{{Measurement of baryon acoustic oscillation
  correlations at z = 2.3 with SDSS DR12 Ly{$\alpha$}-Forests}},
  \href{https://doi.org/10.1051/0004-6361/201730533}{\emph{\aap} {\bfseries
  603} (2017) A12} [\href{https://arxiv.org/abs/1702.00176}{{\ttfamily
  1702.00176}}].

\bibitem{dMdB2020}
H.~{du Mas des Bourboux}, J.~{Rich}, A.~{Font-Ribera}, V.~{de Sainte Agathe},
  J.~{Farr}, T.~{Etourneau} et~al., \emph{{The Completed SDSS-IV Extended
  Baryon Oscillation Spectroscopic Survey: Baryon Acoustic Oscillations with
  Ly{\ensuremath{\alpha}} Forests}},
  \href{https://doi.org/10.3847/1538-4357/abb085}{\emph{\apj} {\bfseries 901}
  (2020) 153} [\href{https://arxiv.org/abs/2007.08995}{{\ttfamily
  2007.08995}}].

\bibitem{DESI2023b.KP1.EDR}
{DESI Collaboration}, A.G.~{Adame}, J.~{Aguilar}, S.~{Ahlen}, S.~{Alam},
  G.~{Aldering} et~al., \emph{{The Early Data Release of the Dark Energy
  Spectroscopic Instrument}},
  \href{https://doi.org/10.48550/arXiv.2306.06308}{\emph{arXiv e-prints} (2023)
  arXiv:2306.06308} [\href{https://arxiv.org/abs/2306.06308}{{\ttfamily
  2306.06308}}].

\bibitem{LyaBAO.EDR.Gordon.2023}
C.~{Gordon}, A.~{Cuceu}, J.~{Chaves-Montero}, A.~{Font-Ribera},
  A.X.~{Gonz{\'a}lez-Morales}, J.~{Aguilar} et~al., \emph{{3D correlations in
  the Lyman-{\ensuremath{\alpha}} forest from early DESI data}},
  \href{https://doi.org/10.1088/1475-7516/2023/11/045}{\emph{\jcap} {\bfseries
  2023} (2023) 045} [\href{https://arxiv.org/abs/2308.10950}{{\ttfamily
  2308.10950}}].

\bibitem{Ramirez2024}
C.~{Ram{\'\i}rez-P{\'e}rez}, I.~{P{\'e}rez-R{\`a}fols}, A.~{Font-Ribera},
  M.A.~{Karim}, E.~{Armengaud}, J.~{Bautista} et~al., \emph{{The
  Lyman-{\ensuremath{\alpha}} forest catalogue from the Dark Energy
  Spectroscopic Instrument Early Data Release}},
  \href{https://doi.org/10.1093/mnras/stad3781}{\emph{\mnras} {\bfseries 528}
  (2024) 6666} [\href{https://arxiv.org/abs/2306.06312}{{\ttfamily
  2306.06312}}].

\bibitem{DESI2024.I.DR1}
{DESI Collaboration}, \emph{{DESI 2024 I: Data Release 1 of the Dark Energy
  Spectroscopic Instrument}}, {\emph{in preparation} (2024) }.

\bibitem{DESI2024.IV.KP6}
{DESI Collaboration}, A.G.~{Adame}, J.~{Aguilar}, S.~{Ahlen}, S.~{Alam},
  D.M.~{Alexander} et~al., \emph{{DESI 2024 IV: Baryon Acoustic Oscillations
  from the Lyman Alpha Forest}},
  \href{https://doi.org/10.48550/arXiv.2404.03001}{\emph{arXiv e-prints} (2024)
  arXiv:2404.03001} [\href{https://arxiv.org/abs/2404.03001}{{\ttfamily
  2404.03001}}].

\bibitem{KP6s6-Cuceu}
A.~{Cuceu}, H.K.~{Herrera-Alcantar}, C.~{Gordon}, P.~{Martini}, J.~{Guy},
  A.~{Font-Ribera} et~al., \emph{{Validation of the DESI 2024 Ly$\alpha$ forest
  BAO analysis using synthetic datasets}},
  \href{https://doi.org/10.48550/arXiv.2404.03004}{\emph{arXiv e-prints} (2024)
  arXiv:2404.03004} [\href{https://arxiv.org/abs/2404.03004}{{\ttfamily
  2404.03004}}].

\bibitem{KP6s5-Guy}
J.~{Guy}, S.G.A.~{Gontcho}, E.~{Armengaud}, A.~{Brodzeller}, A.~{Cuceu},
  A.~{Font-Ribera} et~al., \emph{{Characterization of contaminants in the
  Lyman-alpha forest auto-correlation with DESI}},
  \href{https://doi.org/10.48550/arXiv.2404.03003}{\emph{arXiv e-prints} (2024)
  arXiv:2404.03003} [\href{https://arxiv.org/abs/2404.03003}{{\ttfamily
  2404.03003}}].

\bibitem{Foltz1990}
C.B.~{Foltz}, F.H.~{Chaffee}, P.C.~{Hewett}, R.J.~{Weymann} and S.L.~{Morris},
  \emph{{On the Fraction of Optically-Selected QSOs with Broad Absorption Lines
  in Their Spectra}},  in \emph{Bulletin of the American Astronomical Society},
  vol.~22, p.~806, Mar., 1990.

\bibitem{Trump2006}
J.R.~{Trump}, P.B.~{Hall}, T.A.~{Reichard}, G.T.~{Richards}, D.P.~{Schneider},
  D.E.~{Vanden Berk} et~al., \emph{{A Catalog of Broad Absorption Line Quasars
  from the Sloan Digital Sky Survey Third Data Release}},
  \href{https://doi.org/10.1086/503834}{\emph{\apjs} {\bfseries 165} (2006) 1}
  [\href{https://arxiv.org/abs/astro-ph/0603070}{{\ttfamily
  astro-ph/0603070}}].

\bibitem{Paris2017}
I.~{P{\^a}ris}, P.~{Petitjean}, N.P.~{Ross}, A.D.~{Myers}, {\'E}.~{Aubourg},
  A.~{Streblyanska} et~al., \emph{{The Sloan Digital Sky Survey Quasar Catalog:
  Twelfth data release}},
  \href{https://doi.org/10.1051/0004-6361/201527999}{\emph{\aap} {\bfseries
  597} (2017) A79} [\href{https://arxiv.org/abs/1608.06483}{{\ttfamily
  1608.06483}}].

\bibitem{Filbert2023}
S.~{Filbert}, P.~{Martini}, K.~{Seebaluck}, L.~{Ennesser}, D.M.~{Alexander},
  A.~{Bault} et~al., \emph{{Broad Absorption Line Quasars in the Dark Energy
  Spectroscopic Instrument Early Data Release}},
  \href{https://doi.org/10.48550/arXiv.2309.03434}{\emph{arXiv e-prints} (2023)
  arXiv:2309.03434} [\href{https://arxiv.org/abs/2309.03434}{{\ttfamily
  2309.03434}}].

\bibitem{Ennesser2022}
L.~{Ennesser}, P.~{Martini}, A.~{Font-Ribera} and I.~{P{\'e}rez-R{\`a}fols},
  \emph{{The impact and mitigation of broad-absorption-line quasars in Lyman
  {\ensuremath{\alpha}} forest correlations}},
  \href{https://doi.org/10.1093/mnras/stac301}{\emph{\mnras} {\bfseries 511}
  (2022) 3514} [\href{https://arxiv.org/abs/2111.09439}{{\ttfamily
  2111.09439}}].

\bibitem{DESI2024.II.KP3}
{DESI Collaboration}, \emph{{DESI 2024 II: Two Point Clustering Measurements
  and Validation}}, {\emph{in preparation} (2024) }.

\bibitem{DESI2024.III.KP4}
{DESI Collaboration}, A.G.~{Adame}, J.~{Aguilar}, S.~{Ahlen}, S.~{Alam},
  D.M.~{Alexander} et~al., \emph{{DESI 2024 III: Baryon Acoustic Oscillations
  from Galaxies and Quasars}},
  \href{https://doi.org/10.48550/arXiv.2404.03000}{\emph{arXiv e-prints} (2024)
  arXiv:2404.03000} [\href{https://arxiv.org/abs/2404.03000}{{\ttfamily
  2404.03000}}].

\bibitem{DESI2024.V.KP5}
{DESI Collaboration}, \emph{{DESI 2024 V: Fullshape from Galaxies and
  Quasars}}, {\emph{in preparation} (2024) }.

\bibitem{DESI2024.VI.KP7A}
{DESI Collaboration}, A.G.~{Adame}, J.~{Aguilar}, S.~{Ahlen}, S.~{Alam},
  D.M.~{Alexander} et~al., \emph{{DESI 2024 VI: Cosmological Constraints from
  the Measurements of Baryon Acoustic Oscillations}},
  \href{https://doi.org/10.48550/arXiv.2404.03002}{\emph{arXiv e-prints} (2024)
  arXiv:2404.03002} [\href{https://arxiv.org/abs/2404.03002}{{\ttfamily
  2404.03002}}].

\bibitem{DESI2024.VII.KP7B}
{DESI Collaboration}, \emph{{DESI 2024 VII: Cosmological Constraints from the
  Fullshape Measurements}}, {\emph{in preparation} (2024) }.

\bibitem{DESI2024.VIII.KP7C}
{DESI Collaboration}, \emph{{DESI 2024 VIII: Constraints on Primordial
  Non-Gaussianities}}, {\emph{in preparation} (2024) }.

\bibitem{FocalPlane.Silber.2023}
J.H.~{Silber}, P.~{Fagrelius}, K.~{Fanning}, M.~{Schubnell}, J.N.~{Aguilar},
  S.~{Ahlen} et~al., \emph{{The Robotic Multiobject Focal Plane System of the
  Dark Energy Spectroscopic Instrument (DESI)}},
  \href{https://doi.org/10.3847/1538-3881/ac9ab1}{\emph{\aj} {\bfseries 165}
  (2023) 9} [\href{https://arxiv.org/abs/2205.09014}{{\ttfamily 2205.09014}}].

\bibitem{Corrector.Miller.2023}
T.N.~{Miller}, P.~{Doel}, G.~{Gutierrez}, R.~{Besuner}, D.~{Brooks}, G.~{Gallo}
  et~al., \emph{{The Optical Corrector for the Dark Energy Spectroscopic
  Instrument}}, \href{https://doi.org/10.48550/arXiv.2306.06310}{\emph{arXiv
  e-prints} (2023) arXiv:2306.06310}
  [\href{https://arxiv.org/abs/2306.06310}{{\ttfamily 2306.06310}}].

\bibitem{Fibers.Poppett.2024}
{C. Poppett et al.}, \emph{{The Fiber System for the Dark Energy Spectroscopic
  Instrument}}, {\emph{\apj, submitted} (2024) }.

\bibitem{DESI2022.KP1.Instr}
{DESI Collaboration}, B.~{Abareshi}, J.~{Aguilar}, S.~{Ahlen}, S.~{Alam},
  D.M.~{Alexander} et~al., \emph{{Overview of the Instrumentation for the Dark
  Energy Spectroscopic Instrument}},
  \href{https://doi.org/10.3847/1538-3881/ac882b}{\emph{\aj} {\bfseries 164}
  (2022) 207} [\href{https://arxiv.org/abs/2205.10939}{{\ttfamily
  2205.10939}}].

\bibitem{LS.Overview.Dey.2019}
A.~{Dey}, D.J.~{Schlegel}, D.~{Lang}, R.~{Blum}, K.~{Burleigh}, X.~{Fan}
  et~al., \emph{{Overview of the DESI Legacy Imaging Surveys}},
  \href{https://doi.org/10.3847/1538-3881/ab089d}{\emph{\aj} {\bfseries 157}
  (2019) 168} [\href{https://arxiv.org/abs/1804.08657}{{\ttfamily
  1804.08657}}].

\bibitem{TS.Pipeline.Myers.2023}
A.D.~{Myers}, J.~{Moustakas}, S.~{Bailey}, B.A.~{Weaver}, A.P.~{Cooper},
  J.E.~{Forero-Romero} et~al., \emph{{The Target-selection Pipeline for the
  Dark Energy Spectroscopic Instrument}},
  \href{https://doi.org/10.3847/1538-3881/aca5f9}{\emph{\aj} {\bfseries 165}
  (2023) 50} [\href{https://arxiv.org/abs/2208.08518}{{\ttfamily 2208.08518}}].

\bibitem{Spectro.Pipeline.Guy.2023}
J.~{Guy}, S.~{Bailey}, A.~{Kremin}, S.~{Alam}, D.M.~{Alexander}, C.~{Allende
  Prieto} et~al., \emph{{The Spectroscopic Data Processing Pipeline for the
  Dark Energy Spectroscopic Instrument}},
  \href{https://doi.org/10.3847/1538-3881/acb212}{\emph{\aj} {\bfseries 165}
  (2023) 144} [\href{https://arxiv.org/abs/2209.14482}{{\ttfamily
  2209.14482}}].

\bibitem{SurveyOps.Schlafly.2023}
E.F.~{Schlafly}, D.~{Kirkby}, D.J.~{Schlegel}, A.D.~{Myers}, A.~{Raichoor},
  K.~{Dawson} et~al., \emph{{Survey Operations for the Dark Energy
  Spectroscopic Instrument}},
  \href{https://doi.org/10.3847/1538-3881/ad0832}{\emph{\aj} {\bfseries 166}
  (2023) 259} [\href{https://arxiv.org/abs/2306.06309}{{\ttfamily
  2306.06309}}].

\bibitem{QSOPrelim.Yeche.2020}
C.~{Y{\`e}che}, N.~{Palanque-Delabrouille}, C.-A.~{Claveau}, D.D.~{Brooks},
  E.~{Chaussidon}, T.M.~{Davis} et~al., \emph{{Preliminary Target Selection for
  the DESI Quasar (QSO) Sample}},
  \href{https://doi.org/10.3847/2515-5172/abc01a}{\emph{Research Notes of the
  American Astronomical Society} {\bfseries 4} (2020) 179}
  [\href{https://arxiv.org/abs/2010.11280}{{\ttfamily 2010.11280}}].

\bibitem{DESI2023a.KP1.SV}
{DESI Collaboration}, A.G.~{Adame}, J.~{Aguilar}, S.~{Ahlen}, S.~{Alam},
  G.~{Aldering} et~al., \emph{{Validation of the Scientific Program for the
  Dark Energy Spectroscopic Instrument}},
  \href{https://doi.org/10.3847/1538-3881/ad0b08}{\emph{\aj} {\bfseries 167}
  (2024) 62} [\href{https://arxiv.org/abs/2306.06307}{{\ttfamily 2306.06307}}].

\bibitem{Chaussidon2023}
E.~{Chaussidon}, C.~{Y{\`e}che}, N.~{Palanque-Delabrouille}, D.M.~{Alexander},
  J.~{Yang}, S.~{Ahlen} et~al., \emph{{Target Selection and Validation of DESI
  Quasars}}, \href{https://doi.org/10.3847/1538-4357/acb3c2}{\emph{\apj}
  {\bfseries 944} (2023) 107}
  [\href{https://arxiv.org/abs/2208.08511}{{\ttfamily 2208.08511}}].

\bibitem{VIQSO.Alexander.2023}
D.M.~{Alexander}, T.M.~{Davis}, E.~{Chaussidon}, V.A.~{Fawcett}, A.~{X.
  Gonzalez-Morales}, T.-W.~{Lan} et~al., \emph{{The DESI Survey Validation:
  Results from Visual Inspection of the Quasar Survey Spectra}},
  \href{https://doi.org/10.3847/1538-3881/acacfc}{\emph{\aj} {\bfseries 165}
  (2023) 124} [\href{https://arxiv.org/abs/2208.08517}{{\ttfamily
  2208.08517}}].

\bibitem{Redrock.Bailey.2024}
{S. Bailey et al.}, \emph{{Redrock}}, {\emph{in preparation} (2024) }.

\bibitem{Busca2018}
N.~{Busca} and C.~{Balland}, \emph{{QuasarNET: Human-level spectral
  classification and redshifting with Deep Neural Networks}},
  \href{https://doi.org/10.48550/arXiv.1808.09955}{\emph{arXiv e-prints} (2018)
  arXiv:1808.09955} [\href{https://arxiv.org/abs/1808.09955}{{\ttfamily
  1808.09955}}].

\bibitem{Brodzeller2023}
A.~{Brodzeller}, K.~{Dawson}, S.~{Bailey}, J.~{Yu}, A.J.~{Ross}, A.~{Bault}
  et~al., \emph{{Performance of the Quasar Spectral Templates for the Dark
  Energy Spectroscopic Instrument}},
  \href{https://doi.org/10.3847/1538-3881/ace35d}{\emph{\aj} {\bfseries 166}
  (2023) 66} [\href{https://arxiv.org/abs/2305.10426}{{\ttfamily 2305.10426}}].

\bibitem{KP6s4-Bault}
A.~{Bault}, D.~{Kirkby}, J.~{Guy}, A.~{Brodzeller}, J.~{Aguilar}, S.~{Ahlen}
  et~al., \emph{{Impact of Systematic Redshift Errors on the Cross-correlation
  of the Lyman-$\alpha$ Forest with Quasars at Small Scales Using DESI Early
  Data}}, \href{https://doi.org/10.48550/arXiv.2402.18009}{\emph{arXiv
  e-prints} (2024) arXiv:2402.18009}
  [\href{https://arxiv.org/abs/2402.18009}{{\ttfamily 2402.18009}}].

\bibitem{Herrera2024}
H.K.~{Herrera-Alcantar}, A.~{Mu{\~n}oz-Guti{\'e}rrez}, T.~{Tan},
  A.X.~{Gonz{\'a}lez-Morales}, A.~{Font-Ribera}, J.~{Guy} et~al.,
  \emph{{Synthetic spectra for Lyman-$\alpha$ forest analysis in the Dark
  Energy Spectroscopic Instrument}},
  \href{https://doi.org/10.48550/arXiv.2401.00303}{\emph{arXiv e-prints} (2023)
  arXiv:2401.00303} [\href{https://arxiv.org/abs/2401.00303}{{\ttfamily
  2401.00303}}].

\bibitem{coles91}
P.~{Coles} and B.~{Jones}, \emph{{A lognormal model for the cosmological mass
  distribution.}}, \href{https://doi.org/10.1093/mnras/248.1.1}{\emph{\mnras}
  {\bfseries 248} (1991) 1}.

\bibitem{Farr2020_LyaCoLoRe}
J.~{Farr}, A.~{Font-Ribera}, H.~{du Mas des Bourboux},
  A.~{Mu{\~n}oz-Guti{\'e}rrez}, F.J.~{S{\'a}nchez}, A.~{Pontzen} et~al.,
  \emph{{LyaCoLoRe: synthetic datasets for current and future
  Lyman-{\ensuremath{\alpha}} forest BAO surveys}},
  \href{https://doi.org/10.1088/1475-7516/2020/03/068}{\emph{\jcap} {\bfseries
  2020} (2020) 068} [\href{https://arxiv.org/abs/1912.02763}{{\ttfamily
  1912.02763}}].

\bibitem{Etourneau2023}
T.~{Etourneau}, J.-M.~{Le Goff}, J.~{Rich}, T.~{Tan}, A.~{Cuceu}, S.~{Ahlen}
  et~al., \emph{{Mock data sets for the Eboss and DESI Lyman-$\alpha$ forest
  surveys}}, \href{https://doi.org/10.48550/arXiv.2310.18996}{\emph{arXiv
  e-prints} (2023) arXiv:2310.18996}
  [\href{https://arxiv.org/abs/2310.18996}{{\ttfamily 2310.18996}}].

\bibitem{Ramirez2022}
C.~{Ram{\'\i}rez-P{\'e}rez}, J.~{Sanchez}, D.~{Alonso} and A.~{Font-Ribera},
  \emph{{CoLoRe: fast cosmological realisations over large volumes with
  multiple tracers}},
  \href{https://doi.org/10.1088/1475-7516/2022/05/002}{\emph{\jcap} {\bfseries
  2022} (2022) 002} [\href{https://arxiv.org/abs/2111.05069}{{\ttfamily
  2111.05069}}].

\bibitem{Bi1997}
H.~{Bi} and A.F.~{Davidsen}, \emph{{Evolution of Structure in the Intergalactic
  Medium and the Nature of the Ly{\ensuremath{\alpha}} Forest}},
  \href{https://doi.org/10.1086/303908}{\emph{\apj} {\bfseries 479} (1997) 523}
  [\href{https://arxiv.org/abs/astro-ph/9611062}{{\ttfamily
  astro-ph/9611062}}].

\bibitem{Croft1998}
R.A.C.~{Croft}, D.H.~{Weinberg}, N.~{Katz} and L.~{Hernquist}, \emph{{Recovery
  of the Power Spectrum of Mass Fluctuations from Observations of the
  Ly{\ensuremath{\alpha}} Forest}},
  \href{https://doi.org/10.1086/305289}{\emph{\apj} {\bfseries 495} (1998) 44}
  [\href{https://arxiv.org/abs/astro-ph/9708018}{{\ttfamily
  astro-ph/9708018}}].

\bibitem{Gibson2009}
R.R.~{Gibson}, L.~{Jiang}, W.N.~{Brandt}, P.B.~{Hall}, Y.~{Shen}, J.~{Wu}
  et~al., \emph{{A Catalog of Broad Absorption Line Quasars in Sloan Digital
  Sky Survey Data Release 5}},
  \href{https://doi.org/10.1088/0004-637X/692/1/758}{\emph{\apj} {\bfseries
  692} (2009) 758} [\href{https://arxiv.org/abs/0810.2747}{{\ttfamily
  0810.2747}}].

\bibitem{Niu2020}
W.~{Niu}, \emph{{Better Lyman Alpha Analysis for DESI Cosmology}},  in
  \emph{American Astronomical Society Meeting Abstracts \#235}, vol.~235 of
  \emph{American Astronomical Society Meeting Abstracts}, p.~108.06, Jan.,
  2020.

\bibitem{Guo2019}
Z.~{Guo} and P.~{Martini}, \emph{{Classification of Broad Absorption Line
  Quasars with a Convolutional Neural Network}},
  \href{https://doi.org/10.3847/1538-4357/ab2590}{\emph{\apj} {\bfseries 879}
  (2019) 72} [\href{https://arxiv.org/abs/1901.04506}{{\ttfamily 1901.04506}}].

\bibitem{Weymann1991}
R.J.~Weymann, S.L.~Morris, C.B.~Foltz and P.C.~Hewett, \emph{Comparisons of the
  emission-line and continuum properties of broad absorption line and normal
  quasi-stellar objects},
  \href{https://doi.org/10.1086/170020}{\emph{Astrophysical Journal} {\bfseries
  373} (1991) 23}.

\bibitem{Hall2002}
P.B.~{Hall}, S.F.~{Anderson}, M.A.~{Strauss}, D.G.~{York}, G.T.~{Richards},
  X.~{Fan} et~al., \emph{{Unusual Broad Absorption Line Quasars from the Sloan
  Digital Sky Survey}}, \href{https://doi.org/10.1086/340546}{\emph{\apjs}
  {\bfseries 141} (2002) 267}
  [\href{https://arxiv.org/abs/astro-ph/0203252}{{\ttfamily
  astro-ph/0203252}}].

\bibitem{Leighly2019}
K.M.~{Leighly}, D.M.~{Terndrup}, A.B.~{Lucy}, H.~{Choi}, S.C.~{Gallagher},
  G.T.~{Richards} et~al., \emph{{The z = 0.54 LoBAL Quasar SDSS
  J085053.12+445122.5. II. The Nature of Partial Covering in the
  Broad-absorption-line Outflow}},
  \href{https://doi.org/10.3847/1538-4357/ab212a}{\emph{\apj} {\bfseries 879}
  (2019) 27} [\href{https://arxiv.org/abs/1811.04174}{{\ttfamily 1811.04174}}].

\bibitem{Masribas2019}
L.~Mas-Ribas and R.~Mauland, \emph{The ubiquitous imprint of radiative
  acceleration in the mean absorption spectrum of quasar outflows},
  \href{https://doi.org/10.3847/1538-4357/ab4efd}{\emph{The Astrophysical
  Journal} {\bfseries 886} (2019) 151}.

\bibitem{Lyke2020}
B.W.~{Lyke}, A.N.~{Higley}, J.N.~{McLane}, D.P.~{Schurhammer}, A.D.~{Myers},
  A.J.~{Ross} et~al., \emph{{The Sloan Digital Sky Survey Quasar Catalog:
  Sixteenth Data Release}},
  \href{https://doi.org/10.3847/1538-4365/aba623}{\emph{\apjs} {\bfseries 250}
  (2020) 8} [\href{https://arxiv.org/abs/2007.09001}{{\ttfamily 2007.09001}}].

\bibitem{Brodzeller2022}
A.~{Brodzeller} and K.~{Dawson}, \emph{{Modeling the Spectral Diversity of
  Quasars in the Sixteenth Data Release from the Sloan Digital Sky Survey}},
  \href{https://doi.org/10.3847/1538-3881/ac4600}{\emph{\aj} {\bfseries 163}
  (2022) 110} [\href{https://arxiv.org/abs/2110.07748}{{\ttfamily
  2110.07748}}].

\bibitem{Paris2018}
I.~{P{\^a}ris}, P.~{Petitjean}, {\'E}.~{Aubourg}, A.D.~{Myers},
  A.~{Streblyanska}, B.W.~{Lyke} et~al., \emph{{The Sloan Digital Sky Survey
  Quasar Catalog: Fourteenth data release}},
  \href{https://doi.org/10.1051/0004-6361/201732445}{\emph{\aap} {\bfseries
  613} (2018) A51} [\href{https://arxiv.org/abs/1712.05029}{{\ttfamily
  1712.05029}}].

\bibitem{Garcia2023}
L.{\'A}.~{Garc{\'\i}a}, P.~{Martini}, A.X.~{Gonzalez-Morales},
  A.~{Font-Ribera}, H.K.~{Herrera-Alcantar}, J.N.~{Aguilar} et~al.,
  \emph{{Analysis of the impact of broad absorption lines on quasar redshift
  measurements with synthetic observations}},
  \href{https://doi.org/10.1093/mnras/stad2993}{\emph{\mnras} {\bfseries 526}
  (2023) 4848} [\href{https://arxiv.org/abs/2304.05855}{{\ttfamily
  2304.05855}}].

\bibitem{DESI_SV}
A.G.~{Adame}, J.~{Aguilar}, S.~{Ahlen}, S.~{Alam}, G.~{Aldering},
  D.M.~{Alexander} et~al., \emph{{Validation of the Scientific Program for the
  Dark Energy Spectroscopic Instrument}},
  \href{https://doi.org/10.3847/1538-3881/ad0b08}{\emph{\aj} {\bfseries 167}
  (2024) 62} [\href{https://arxiv.org/abs/2306.06307}{{\ttfamily 2306.06307}}].

\bibitem{Kaiser1987}
N.~{Kaiser}, \emph{{Clustering in real space and in redshift space}},
  \href{https://doi.org/10.1093/mnras/227.1.1}{\emph{\mnras} {\bfseries 227}
  (1987) 1}.

\bibitem{Cuceu2021}
A.~{Cuceu}, A.~{Font-Ribera}, B.~{Joachimi} and S.~{Nadathur}, \emph{{Cosmology
  beyond BAO from the 3D distribution of the Lyman-{\ensuremath{\alpha}}
  forest}}, \href{https://doi.org/10.1093/mnras/stab1999}{\emph{\mnras}
  {\bfseries 506} (2021) 5439}
  [\href{https://arxiv.org/abs/2103.14075}{{\ttfamily 2103.14075}}].

\bibitem{Cuceu2023a}
A.~{Cuceu}, A.~{Font-Ribera}, S.~{Nadathur}, B.~{Joachimi} and P.~{Martini},
  \emph{{Constraints on the Cosmic Expansion Rate at Redshift 2.3 from the
  Lyman-{\ensuremath{\alpha}} Forest}},
  \href{https://doi.org/10.1103/PhysRevLett.130.191003}{\emph{\prl} {\bfseries
  130} (2023) 191003} [\href{https://arxiv.org/abs/2209.13942}{{\ttfamily
  2209.13942}}].

\bibitem{Cuceu2023b}
A.~{Cuceu}, A.~{Font-Ribera}, P.~{Martini}, B.~{Joachimi}, S.~{Nadathur},
  J.~{Rich} et~al., \emph{{The Alcock-Paczy{\'n}ski effect from
  Lyman-{\ensuremath{\alpha}} forest correlations: analysis validation with
  synthetic data}}, \href{https://doi.org/10.1093/mnras/stad1546}{\emph{\mnras}
  {\bfseries 523} (2023) 3773}
  [\href{https://arxiv.org/abs/2209.12931}{{\ttfamily 2209.12931}}].

\bibitem{Karacayli2024}
N.G.~{Kara{\c{c}}ayl{\i}}, P.~{Martini}, J.~{Guy}, C.~{Ravoux}, M.L.A.~{Karim},
  E.~{Armengaud} et~al., \emph{{Optimal 1D Ly{\ensuremath{\alpha}} Forest Power
  Spectrum Estimation - III. DESI early data}},
  \href{https://doi.org/10.1093/mnras/stae171}{\emph{\mnras} (2024) }
  [\href{https://arxiv.org/abs/2306.06316}{{\ttfamily 2306.06316}}].

\bibitem{Ravoux2023}
C.~{Ravoux}, M.L.~{Abdul Karim}, E.~{Armengaud}, M.~{Walther},
  N.G.~{Kara{\c{c}}ayl{\i}}, P.~{Martini} et~al., \emph{{The Dark Energy
  Spectroscopic Instrument: one-dimensional power spectrum from first Ly
  {\ensuremath{\alpha}} forest samples with Fast Fourier Transform}},
  \href{https://doi.org/10.1093/mnras/stad3008}{\emph{\mnras} {\bfseries 526}
  (2023) 5118} [\href{https://arxiv.org/abs/2306.06311}{{\ttfamily
  2306.06311}}].

\end{thebibliography}\endgroup

\end{document}